\documentclass[lettersize,journal]{IEEEtran}
\usepackage{algorithm}
\usepackage{algorithmicx}
\usepackage{algpseudocode}
\usepackage{amsmath,amsfonts,amssymb,amsthm}
\usepackage{array}
\usepackage{subfig}
\usepackage{textcomp}
\usepackage{stfloats}
\usepackage{url}
\usepackage{verbatim}
\usepackage{graphicx}
\usepackage{bm}
\usepackage{svg}
\usepackage{cite}
\usepackage{threeparttable}
\usepackage{multirow}
\usepackage{lettrine}
\usepackage[colorlinks=true, linkcolor=blue, citecolor=blue, urlcolor=blue]{hyperref}
\usepackage{mathrsfs}
\usepackage{booktabs} 
\usepackage[labelsep=period]{caption}
\usepackage{mathtools}

\makeatletter

\newcommand{\Rmnum}[1]{\expandafter@slowromancap\romannumeral #1@}
\makeatother
\theoremstyle{remark}

\begin{document} 

\title{An Asynchronous Triggered MAC Protocol for Underwater Acoustic Networks}

\author{Bingwen Huangfu,
  Jiani Guo, 
  Shanshan Song*,~\IEEEmembership{Member, IEEE},
  Nan Sun,
  Jun Liu,
  and Miao Pan,~\IEEEmembership{Senior Member, IEEE}
  \thanks{
    S. Song is the corresponding author.}
  \thanks{
  B. Huangfu, J. Guo, and S. Song are with the College of Computer Science and Technology, Jilin University, Changchun 130012, China (e-mail: hfbw24@mails.jlu.edu.cn, jnguo@jlu.edu.cn, songss@jlu.edu.cn).
  }
\thanks{
  N. Sun is with the College of Software Engineering, Jilin University, Changchun 130012, China (e-mail: nsun23@mails.jlu.edu.cn).
  }
  \thanks{
    Jun Liu is with the School of Electronic and Information Engineering, Beihang University, Beijing 100191, China (e-mail: liujun2019@buaa.edu.cn).
    }
  \thanks{
  Miao Pan is with the Department of Electrical and Computer Engineering, University of Houston, Houston, TX 77204 USA (e-mail: mpan2@uh.edu).  
}
}

\maketitle
\begin{abstract}

  Time Division Multiple Access (TDMA)-based Medium Access Control (MAC) protocols have proven their practicality through extensive field trials in Underwater Acoustic Networks (UANs), attributable to their hardware compatibility and ease of implementation. 
  In conventional TDMA-based MAC designs, channel access is typically organized using synchronized, fixed-length slots to mitigate contention and coordinate transmissions.
  However, this paradigm imposes significant clock synchronization overhead in UANs with long and variable propagation delays and struggles to improve scheduling flexibility. 
  Although some protocols attempt to refine this slot paradigm (adjust the slot length to improve channel reuse efficiency or scheduling frequency), they are still constrained by the trade-off between channel utilization and scheduling complexity.
  To this end, this paper proposes AT-MAC, an \underbar{A}synchronous \underbar{T}riggered MAC protocol that aims to achieve efficient and fair channel access through coordinated asynchronous scheduling.
  AT-MAC introduces a triggered slot paradigm without time synchronization, decoupling transmission scheduling from a rigid timeline and enabling asynchronous, variable-length slots to accommodate the long and diverse propagation delays.
  To power this slot paradigm, AT-MAC augments conventional Multi-Agent Deep Reinforcement Learning to handle asynchronous interaction, achieving coordinated channel access under partial observations.
  It further devises a load-aware fairness guard mechanism to enable network-wide fairness status inference solely through local overhearing, thereby guiding adaptive scheduling correction to maintain fairness.
  Field-reconstructed simulations and on-board inference benchmarking demonstrate the feasibility of AT-MAC.  
  Extensive simulation results further demonstrate its consistent performance gains across the evaluated scenarios and traffic conditions.

\end{abstract}
  
  
\section{Introduction}

\IEEEPARstart{U}{nderwater} acoustic networks (UANs) have attracted significant research interest in recent decades due to their applications in ocean exploration, environmental monitoring, disaster prevention, and military surveillance \cite{sozer2000underwater,adhoc,towards,IOUT}.
However, these applications hinge on efficient and reliable data transmission.
To achieve this, medium access control (MAC) protocols play a pivotal role in enhancing network throughput, orchestrating competing channel access, and managing multi-user fairness \cite{DOTS,MCMAC,ACH}.
Specifically, the unique characteristics of underwater acoustic communication, such as long propagation delay, high attenuation, narrow bandwidth, and half-duplex communication, impose formidable challenges to the MAC protocol design in UANs \cite{uchannel}.

Due to the high compatibility with communication schemes and device models, Time Division Multiple Access (TDMA)-based MAC protocols have been extensively adopted in underwater networks across various fields.
Numerous offshore field trials indicate their practicality and ease of implementation \cite{survey,propagation,2013Comparing,pmac}.
Conventional TDMA-based MAC protocols rely on synchronized and fixed-length slots to enable nodes to access the channel in an orderly manner and avoid packet conflicts.
However, the effectiveness of this synchronized slot paradigm can be affected by clock synchronization errors \cite{2006Time,PTA-Sync}.
Residual synchronization errors in underwater time synchronization methods lead to persistent clock offsets among nodes that accumulate over time.
Moreover, maintaining time synchronization generally requires extra information exchange, which incurs control overhead in terms of channel and energy within resource-constrained UANs.

Even with ideal clock synchronization, improving channel utilization under a synchronized-slot paradigm often comes at the cost of increased scheduling complexity.
1) Dedicated slot allocation can readily avoid packet collisions but leaves substantial channel idleness, as propagation delays dominate the round-trip latency \cite{sfama,terminal}.
Although exploiting differences in propagation delays allows for spatial slot reuse, its rigid topology-dependence renders transmission concurrency vulnerable \cite{distance,stump,TDA,propagation-delay,Delay,state,park2019reinforcement}.
2) Shortening the slot length improves scheduling frequency, holding promise for dense and staggered channel access.
However, this design may introduce across-slot packet collisions.
This necessitates MAC protocols to predict future channel usage, significantly escalating scheduling complexity \cite{subslot,ye2019deep,deep2,exploiting}.

Furthermore, the synchronized and fixed-length slot paradigm makes it difficult to realize dynamic scheduling to accommodate varying traffic load conditions while ensuring high channel utilization \cite{UIOT,gussen2016survey}.
To introduce the necessary flexibility and efficiency, the integration of shortened time slots and adaptive scheduling methods like Deep Reinforcement Learning (DRL) may become a promising solution.
However, the resulting scheduling complexity makes effective network-wide coordination challenging, particularly when channel utilization and fairness need to be jointly considered.

Motivated by the aforementioned limitations of network-wide synchronized time slots, we propose AT-MAC, an \underbar{A}synchronous \underbar{T}riggered \underbar{MAC} protocol that aims to achieve efficient and fair channel access through coordinated asynchronous scheduling.
Specifically, AT-MAC introduces a triggered slot paradigm without time synchronization, decoupling transmission scheduling from a rigid timeline and supporting flexible asynchronous scheduling.
Based on this paradigm, AT-MAC develops a time-offset asynchronous Multi-Agent Proximal Policy Optimization (MAPPO) algorithm to coordinate asynchronous channel access under partial observations.
A load-aware fairness guard mechanism is further incorporated to regulate transmission decisions and maintain network-wide fairness.

The main contributions of this paper are as follows:
\begin{itemize}

  \item {Design a triggered slot paradigm to enable asynchronous transmission scheduling without clock and slot synchronization.
  All nodes locally maintain slots, each dedicated to a single, complete \textit{DATA-ACK} transaction.
  Asynchronous, variable-length slots provide nodes with the scheduling flexibility to exploit long and diverse propagation delays for naturally staggered channel access.
  }
  \item {Propose an asynchronous time-offset MAPPO algorithm for the triggered slot paradigm.
  We formulate asynchronous channel access as a time-offset decentralized partially observable Markov decision process (To-Dec-POMDP), and introduce relative time offsets for temporally consistent  observations and time-aware discounting for non-uniform interaction intervals.
  Integrated with centralized training and decentralized execution, the algorithm achieves stable convergence and efficient execution.
  }
  \item {Present a low-overhead load-aware fairness guard mechanism that decouples fairness regulation from MAPPO-based channel-access learning, thereby reducing the multi-objective learning burden while maintaining network-wide fairness.
  By leveraging the broadcast nature of acoustic communication, each node can estimate the network-wide load status from locally overheard information and accordingly correct its transmission behavior toward the desired fairness level.
  }
  \item {
Conduct field-reconstructed simulations, on-board inference benchmarking, and extensive performance and ablation studies to evaluate AT-MAC.
The results demonstrate its computational feasibility on embedded underwater hardware and consistent performance gains across diverse topologies and traffic conditions.
Ablation and clock-desynchronization experiments further validate the effectiveness of the fairness guard mechanism and the robustness of AT-MAC under clock desynchronization.
}

\end{itemize}

\section{Related Work}

\subsection{Conventional TDMA-based MAC: Trading Overhead for Efficiency}

To simplify implementation and mitigate collisions, most TDMA-based MAC protocols adopt the network-wide synchronized time slots that encompass a full transmission-feedback cycle.
The authors in \cite{sfama} proposed a protocol to avoid packet collisions by extending the slot duration to encompass the maximum propagation delay plus the transmission time of a control packet, thereby ensuring transmission isolation between slots.
\cite{pmac} adopted the same time slot setting but allowed nodes far apart to transmit data in the same slot to accommodate chain topology, thereby increasing network throughput.
However, the dedicated synchronized slot design inevitably leads to channel wastage due to the long propagation delays of underwater acoustic signals.

To reduce the channel wastage of long synchronized slots, later protocols exploit topology knowledge or handshaking to enable spatial reuse \cite{DOTS,distance,stump,TDA,propagation-delay,Delay}. 
Although effective in relatively static settings, these methods require collecting propagation or scheduling information through repeated exchanges, which is costly in large-scale and resource-constrained UANs. 
Moreover, the feasibility of slot reuse is inherently bounded by network topology, and modifying static scheduling schemes in dynamic environments entails sustained control overhead.

\subsection{RL-based MAC Suffering from Synchronized Slot Limitations}

Reinforcement Learning (RL) has emerged as a promising framework to address the overhead-efficiency trade-off by enabling nodes to learn effective transmission policies through local interaction with the environment. 
The work in \cite{park2019reinforcement} designed an RL-based protocol where all nodes in the network use basic Q-learning to independently find the most appropriate time slots to transmit, using the number of time slots per frame as the state space and updating the Q-values directly.
\cite{park2020reinforcement,DR-ALOHA-Q,PFAloha} adopted the same slot design as \cite{park2019reinforcement} and incorporated additional considerations to enhance adaptability.
The authors in \cite{subslot} designed a probability-based MAC protocol, which optimizes the channel access probability of nodes within the next adjustable time slots to minimize collisions.

All these RL-based MAC protocols enhance channel utilization by reusing either a single long-duration slot or a frame comprising multiple short slots.
While the flexible transmission timing selection within a slot can gain additional opportunities for spatial reuse, its feasibility is dependent on the network topology.
To make such reuse feasible, the slot duration must be sufficiently long, which further exacerbates channel underutilization when network traffic is insufficient.

To further adaptively exploit the substantial unoccupied channel resources, several studies have proposed shortening the time slot duration to conduct fine-grained scheduling, aiming to achieve dense and staggered receptions at the sink node.
\cite{ye2019deep,deep2,exploiting} shorten the time slot length to the duration of data plus ACK transmission, and employ a DRL algorithm to enable nodes to autonomously decide whether to transmit packets based on local observations.
\cite{shi2025delay} proposed a delay-fluctuation-resistant MAC protocol based on Double DQN in the shortened-slot paradigm, which can adapt to the delay fluctuation to avoid the credit assignment bias and achieve better performance.
\cite{jiang2025underwater} proposed a MAPPO-based MAC protocol that allows multiple intelligent nodes to cooperatively access the channel in the shortened-slot paradigm, thereby improving network efficiency and achieving better network fairness.
Such an approach breaks the transmission isolation between slots, significantly enhancing the flexibility of channel access scheduling, enabling intelligent nodes to exploit idle channel.
\cite{delayaware} proposed DA-MARLA, which leverages MAPPO to facilitate cooperative channel access among multiple nodes, thereby improving network efficiency.
DA-MARLA adopts a more radical design by setting the time slot duration equal to the data transmission time and omitting the use of ACKs.
This design avoids the \textit{action time-dependence} caused by waiting for ACKs, reducing scheduling complexity and enabling cooperative channel access among multiple intelligent nodes.

However, these DRL-based MAC protocols with shortened time slots may not fully account for the heterogeneity of propagation delays, potentially leading to issues such as credit assignment bias and self-action masking.
This leads to their reliance on idealized environment assumptions, making it difficult to achieve stable learning and efficient execution in real acoustic propagation environments.

\subsection{Traffic Load Awareness and Fairness in UANs}

Recently, with the increasing number of nodes and the development of underwater applications, the fairness issue in UANs has become more prominent.
Some research has begun to focus on how to achieve load awareness and fair scheduling in UANs.
\cite{hsu2009st} proposed a heuristic link scheduling algorithm (TOTA) that considers traffic load and routing information to save energy and improve throughput. The base station collects the global information such as topology, traffic load on each link, to compute the optimal schedule.
\cite{lt} allowed nodes to attach load information to control packets, achieving traffic-load adaptation by transmitting variable length packets before the scheduled time slots.
\cite{zhang2019load} proposed a load-based time slot allocation scheme to mitigate intra-network interference and improve channel utilization under fluctuating traffic loads.
\cite{su2021traffic} proposed a traffic load-aware link scheduling scheme for UWSNs, where traffic load is evaluated based on the proportion of time the control channel remains busy during the observation period. Distributed nodes exchange control packets to update their channel resource lists.

These strategies necessitate information exchanges between nodes to acquire either global or local traffic load information.
However, such information exchange incurs additional overhead in terms of channel and energy consumption, which is detrimental in resource-constrained UANs.
Therefore, a practical fairness mechanism should maintain network-wide fairness with minimal additional overhead.

\begin{figure}[t]
  \centerline{\includegraphics[width=0.9\linewidth]{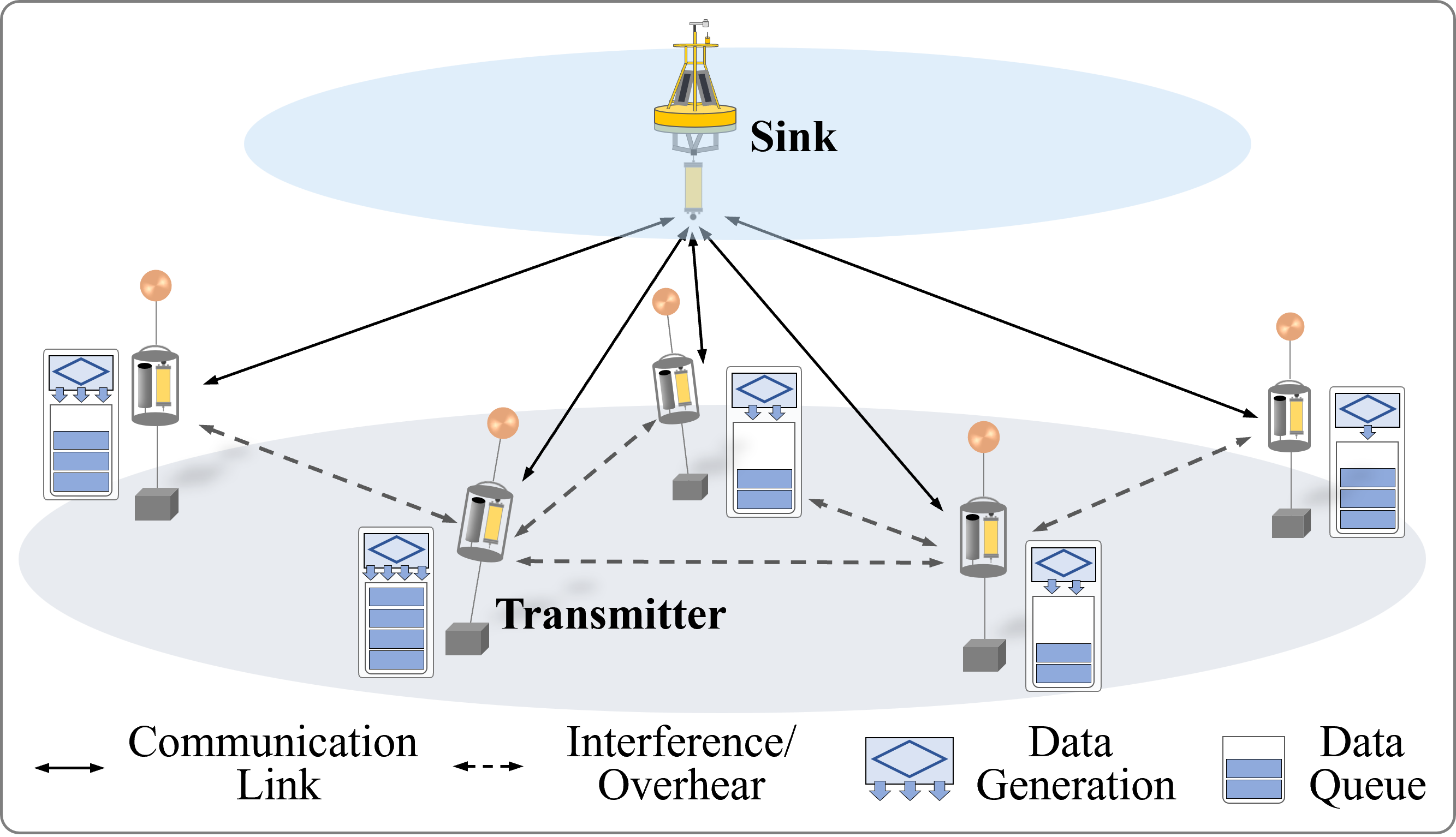}}
  \caption{Overview of the considered system. Transmitters send data packets to a central sink, which returns ACK packets.}
  \label{system}
\end{figure}

\section{System Model and Problem Formulation}

\subsection{System Model}

As shown in Fig.~\ref{system}, we consider a one-hop underwater acoustic network (UAN) where $N$ transmitters send data packets to a sink node over a shared acoustic channel.
Such a topology is typical for small-scale networks and serves as a fundamental building block for large UANs.
We denote the set of transmitters as $\mathcal{N} = \{1,2,\ldots ,N\}$, with the sink node represented as $m$.
Each transmitter $i$ maintains a data queue with length $q_i$ and generates data at various rates $\lambda_i$. 
The transmit power $p_i$ is pre-configured based on the link distance and channel conditions, and includes an adequate margin to ensure reliable connectivity to the sink.
All nodes operate in half-duplex mode, defined as 
$\mathrm{Mode}_i(t) \in \{\mathrm{Tx},\mathrm{Rx},\mathrm{Idle}\}$,
where the tri-mode definition is mutually exclusive, enforcing half-duplex operation such that a node cannot transmit and receive at the same time.

Consider an executed data transmission from transmitter $i$ starting at time $t$.
The complete process consists of four phases:
\begin{enumerate}
\item \textit{Data transmission}: at time $t$, transmitter $i$ sends a data packet to sink $m$ with a transmission delay of $\delta^{\mathrm{tx}}_{l_i} = \delta^{\mathrm{pre}} + l_i/R$, where $\delta^{\mathrm{pre}}$ is the preamble duration and $R$ is the physical-layer transmission rate;
\item \textit{Data propagation and reception}: the packet propagates over delay $\delta^{\mathrm{prop}}_{i,m} = d_{i,m}/c$, where $d_{i,m}$ is the distance between transmitter $i$ and sink $m$, $c$ is the sound speed. 
Sink $m$ receives the packet at time $t + \delta^{\mathrm{tx}}_{l_i} + \delta^{\mathrm{prop}}_{i,m}$;
\item \textit{ACK transmission}: upon successfully receiving the data, sink $m$ sends back an ACK packet with a transmission delay of $\delta^{\mathrm{tx}}_{\mathrm{ack}} = \delta^{\mathrm{pre}}$, since the acknowledgment information can be embedded in the preamble;
\item \textit{ACK propagation and reception}: the ACK packet propagates back to transmitter $i$ with the same one-way delay $\delta^{\mathrm{prop}}_{i,m}$, and is received at time $t' = t + \delta^{\mathrm{tx}}_{l_i} + 2\delta^{\mathrm{prop}}_{i,m} + \delta^{\mathrm{tx}}_{\mathrm{ack}}$.
\end{enumerate}
Upon receiving the ACK, transmitter $i$ considers the data packet delivered and removes it from its queue.

Therefore, we can define the data reception window $W_i^{\mathrm D}$ at the sink and the ACK reception window $W_i^{\mathrm A}$ at transmitter $i$ as:
\begin{equation}
  W_i^{\mathrm D}(t)=\left[t+\delta^{\mathrm{prop}}_{i,m},\ t+\delta^{\mathrm{tx}}_{l_i}+\delta^{\mathrm{prop}}_{i,m}\right],
\end{equation}
\begin{equation}
W_i^{\mathrm A}(t)=\left[t+\delta^{\mathrm{tx}}_{l_i}+2\delta^{\mathrm{prop}}_{i,m},\ t+\delta^{\mathrm{tx}}_{l_i}+2\delta^{\mathrm{prop}}_{i,m}+\delta^{\mathrm{tx}}_{\mathrm{ack}}\right].
\end{equation}
Constrained by the half-duplex operation and the shared narrowband broadcast channel, the transmission of the data packet $\textit{pkt}$ from transmitter $i$ to sink $m$ fails if any of the following events occurs:
\begin{enumerate}
  \item \textit{Data reception collision (at sink).} Another packet overlaps with $W_i^{\mathrm D}(t)$ at the sink, causing destructive interference and corrupting $\textit{pkt}$;

  \item \textit{ACK--data half-duplex collision (at sink).} The sink is transmitting another ACK during $W_i^{\mathrm D}(t)$; its half-duplex operation precludes simultaneous reception of $\textit{pkt}$;

  \item \textit{ACK reception collision (at transmitter).} Another packet arrival overlaps with $W_i^{\mathrm A}(t)$ at transmitter $i$, corrupting the ACK decoding.
\end{enumerate}

Based on the above collision analysis, we define $f_i(t) \in \{-1, 0, 1\}$ as the feedback outcome for transmitter $i$'s decision at time $t$.
Specifically, $f_i = 1$ indicates that the packet is successfully delivered (an ACK is received), $f_i = -1$ indicates a failed attempt (no ACK is received for any reason), and $f_i = 0$ indicates that no transmission is executed.
Table~\ref{notation} summarizes the notations.

\begin{table}[t]
  \centering
  \caption{Notation Description}
  \renewcommand{\arraystretch}{1.1}
  \setlength{\tabcolsep}{4pt}
  \begin{tabular}{c@{\hspace{2pt}}l c@{\hspace{2pt}}l}
    \hline
    \textbf{Notation} & \textbf{Description} & \textbf{Notation} & \textbf{Description} \\
    \hline
    $\mathcal{N}$ & Transmitters set & $N$ & Number of transmitters \\
    $m$ & Sink node & $q_i$ & Data queue length \\
    $\lambda_i$ & Data generation rate & $u_i$ & Transmission decision \\
    $l_i$ & Packet size & $\delta^{\mathrm{pre}}$ & Preamble duration \\
    $\delta^{\mathrm{tx}}_{l_i}$ & Data transmission delay & $\delta^{\mathrm{prop}}_{i,j}$ & Propagation delay \\
    $d_{i,m}$ & Euclidean distance & $c$ & Sound speed \\
    $\delta^{\mathrm{tx}}_{\mathrm{ack}}$ & ACK transmission delay & $f_i$ & Transmission feedback \\
    $l_{\max}$ & Maximum packet size &   $L$ & Observation length \\ 
     $h$ & Fairness horizon & $\epsilon$ & Fairness deviation tolerance \\
    $B^{\mathrm{avail}}_i$ & Available data amount & $B^{\mathrm{deliv}}_i$ & Delivered data amount \\
    \hline
  \end{tabular}
  \label{notation}
\end{table}

\subsection{Problem Formulation}

This paper aims to: 
(i) maximize effective channel utilization by increasing successful receptions and avoiding collision-induced waste;
and (ii) maintain load-aware fairness so that each transmitter obtains transmission opportunities proportional to its load.
However, higher channel utilization requires more concurrent transmissions to exploit spatial reuse, and the feasibility of such concurrency is strongly shaped by topology-induced interference and heterogeneous propagation delays among transmitters.
Specifically, aggressively maximizing channel utilization may enable a subset of transmitters with favorable concurrent transmission capability to dominate channel access over time, thereby causing unfairness.
Given the inherent conflict between these two objectives, we formulate a joint optimization problem to enhance effective channel utilization while ensuring load-aware fairness.

\subsubsection{Effective Utilization}

We define the effective utilization utility ${U}_i(t)$ to quantify the effective channel utilization contributed by a single transmission scheduling action of transmitter $i$ at time $t$, which captures both throughput and collision:
\begin{equation}
    {U}_i(t) = f_i(t)\cdot l_i(t)/l_{\max},
    \label{utilization}
\end{equation}
where $f_i(t)$ is the feedback for the transmission decision at time $t$, and $l_{\max}$ is the maximum allowable packet size.
Given the per-action effective utilization utility $U_i(t)$, the long-term effective utilization of policy $\pi$ is defined as
\begin{equation}
  J_U(\pi)=\mathbb{E}_{\pi}\!\left[\sum_{i\in\mathcal{N}}\sum_{t_i^k \le T} U_i\!\left(t_i^k\right)\right],
  \label{eq:JU}
\end{equation}
where $\pi$ is the transmission scheduling policy, $T$ is the network lifetime, and $t_i^k$ is the $k$-th decision epoch of transmitter $i$.

\subsubsection{Load-Aware Fairness}
Let $h$ denote the fairness horizon. At time $t$, we define the sent-to-available ratio $\rho_i(t)$ of transmitter $i$ over $[t-h,\,t]$ to quantify the proportion of successfully delivered data to the total available data within the horizon $h$, calculated as:
\begin{equation}
  \rho_i(t) = 
  \frac{\sum_{t_i^k \in [t-h,\,t]} \mathbb{I}\{f_i(t_i^k)=1\} \cdot l_i(t_i^k)}
  {q_i(t-h)+\int_{t-h}^{t}\lambda_i(\tau)\,d\tau}. 
  \label{eq:rho}
\end{equation}
Transmitters with no available data within the fairness horizon are excluded from the fairness calculation at time $t$.
Thus, we quantify the network-wide load-aware fairness at time $t$ using Jain's index over $\{\rho_i(t)\}_{i\in\mathcal{N}}$:
\begin{equation}
  F(t)=
    \frac{\left(\sum_{i\in\mathcal{N}}\rho_i(t)\right)^2}
    {N\cdot \sum_{i\in\mathcal{N}}(\rho_i(t))^2}. 
    \label{eq:fairness}
\end{equation}
By definition, $F(t)\in[0,1]$, and $F(t)=1$ when all transmitters exhibit identical $\rho_i(t)$.
The long-term fairness objective is
\begin{equation}
J_F(\pi)=\mathbb{E}_{\pi}\!\left[\int_{0}^{T} F(t)\,dt\right].
\label{eq:JF}
\end{equation}

\subsubsection{Joint Optimization Problem}
Combining the effective utilization and load-aware fairness, we formulate:
\begin{subequations}\label{eq:problem_mo}
  \begin{align}
    \max_{\pi}\quad & \Big(J_U(\pi),\,J_F(\pi)\Big) \label{eq:problem_mo_obj}\\
    \text{s.t.}\quad 
    & u_i(t)\in\{0,1\},\quad \forall i\in\mathcal{N},\ \forall t\in[0,T],\label{eq:const_u}\\
    & 0\le l_i(t)\le \min\{l_{\max},\,q_i(t)\},\quad \forall i\in\mathcal{N},\ \forall t\in[0,T], \label{eq:const_pkt}
  \end{align}
\end{subequations}
where $\pi$ is the transmission scheduling policy, and $T$ is the network lifetime.
Constraint \eqref{eq:const_u} restricts the binary transmission decision of each transmitter.
Constraint \eqref{eq:const_pkt} ensures that the packet size does not exceed the maximum allowable size or the current queue length.

\begin{figure*}
  \centerline{\includegraphics[width=0.85\textwidth]{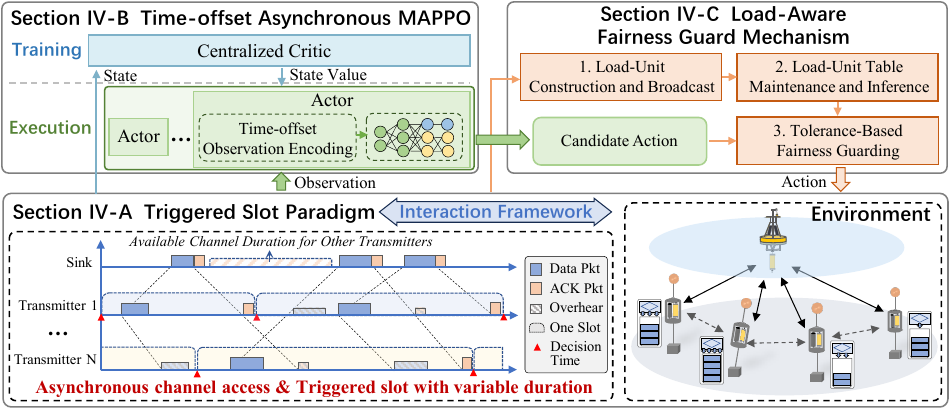}}
  \caption{Overview of the AT-MAC protocol. Built upon the triggered slot paradigm, AT-MAC employs a time-offset asynchronous MAPPO to achieve coordinated scheduling for improved channel utilization, and incorporates a load-aware fairness guard mechanism that corrects transmission decisions to maintain network-wide fairness.}
  \label{flow}
\end{figure*}

\section{AT-MAC Protocol}

To address the joint utilization--fairness formulation in Section~III, we propose AT-MAC, an asynchronous triggered MAC protocol that improves channel utilization while maintaining fairness, without requiring time synchronization. 
As illustrated in Fig.~\ref{flow}, AT-MAC comprises three components: (i) a triggered slot paradigm that defines the protocol's fundamental interaction framework;
(ii) a time-offset asynchronous MAPPO that enables coordinated channel access based on local observations to maximize effective utilization;
and (iii) a load-aware fairness guard mechanism that suppresses transmission decisions violating the fairness tolerance.

AT-MAC decouples fairness from the RL objective to reduce the learning burden associated with jointly optimizing channel utilization and fairness in long-delay underwater acoustic environments.
Within this architecture, AT-MAC leverages MAPPO to prioritize opportunistic utilization subject to the lightweight deterministic fairness constraint, enabling the transmitters to exploit the released channel opportunities resulting from the fairness guard mechanism.

Sections~\ref{sec:tsp}--\ref{sec:fair} detail each component, and Section~\ref{sec:workflow} describes the execution and training workflow.

\begin{figure}[t]
  \captionsetup[subfigure]{font=small} 
  \captionsetup[subfloat]{captionskip=1pt} 
  \centering  
  \subfloat[$u=0$ \& $f=0 $ \label{tts1}]{
      \includegraphics[width=0.98\linewidth]{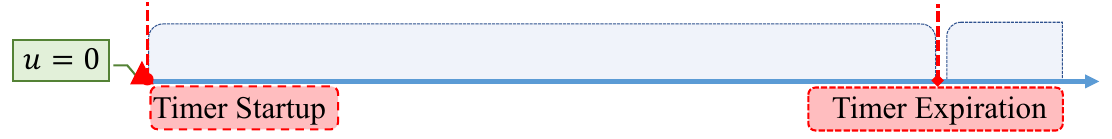}}\\
  \subfloat[$u=1$ \& $f=1$ \label{tts2}]{
      \includegraphics[width=0.98\linewidth]{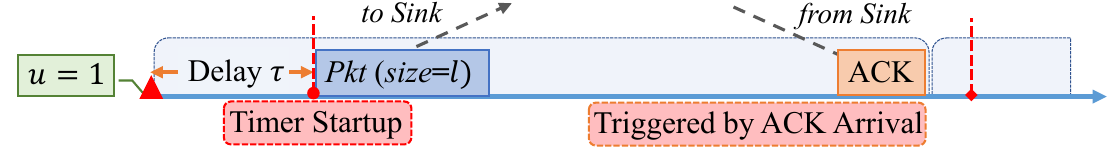}}\\
  \subfloat[$u=1$ \& $f=-1$ \label{tts3}]{
      \includegraphics[width=0.98\linewidth]{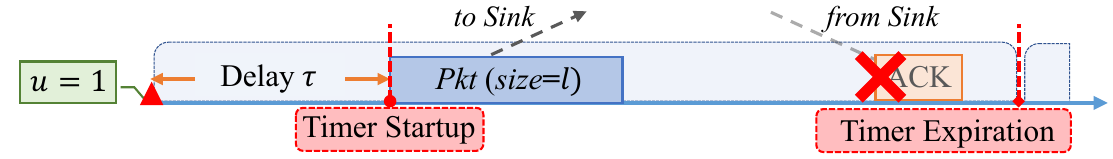}}\\[3pt]
  \includegraphics[width=0.95\linewidth]{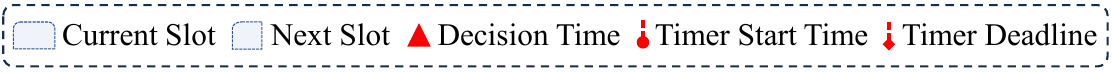}
  \caption{Three state transitions of triggered slot paradigm: (a) no transmission; the transmitter remains idle until timer expiration and enters the next slot; (b) transmission with an ACK received before timer expiration, then enters the next slot; (c) transmission without an ACK, entering the next slot upon timer expiration.
  }
  \label{tts}
\end{figure}

\subsection{Triggered Slot Paradigm \label{sec:tsp}}

Triggered slot paradigm is proposed to provide the basis for asynchronous channel access without requiring time synchronization, where transmitters can independently determine when and how to occupy the acoustic channel.

At the initial stage of the network, transmitter $i$ can make a decision $a_i(t)$ at time $t$, which marks the beginning of a triggered slot.
The decision $a_i(t)$ is a tuple containing three elements, represented as:
\begin{equation}
  a_i(t) = \big(u_i,\, \tau_i,\, l_i\big),
  \label{decision}
\end{equation}
where $u_i\in\{0,1\}$ indicates whether transmitter $i$ chooses to transmit at time $t$, 
$\tau_i$ and $l_i$ are only valid when $u_i=1$, representing that transmitter $i$ sends a data packet of size $l_i$ at time $t + \tau_i$.
To trade off the transmitter's decision flexibility and action space complexity, we define the selectable range of $\tau_i$ as $[0, \tau_{\max}]$, where $\tau_{\max}$ is given by:
\begin{equation}
  \tau_{\max}  = \delta^{\mathrm{tx}}_{l_{\max}}+\delta^{\mathrm{tx}}_{\mathrm{ack}}+ 2 \cdot \frac{d_{\max}}{c},
  \label{delay}
\end{equation}
which represents the worst-case round-trip delay for transmitting a maximum-size packet. 
This includes the transmission delays for a maximum-sized data packet ($l_{\max}$) and its corresponding ACK, plus the two-way propagation delay over the maximum range $d_{\max}$ at a sound speed of $c$.

The fairness guard mechanism maps $a_i(t)$ to the executed action $a_i^{\mathrm{exe}}(t) = (u_i^{\mathrm{exe}}(t),\, \tau_i(t),\, l_i(t))$, as detailed later in Section~\ref{sec:fair}. The executed action and feedback $f_i$ then determine the triggered-slot transition.

\subsubsection{Non-Transmission ($u_i^{\mathrm{exe}}=0$)}
As illustrated in Fig.~\ref{tts}\subref{tts1}, if transmitter $i$ executes non-transmission at time $t$, it starts a timer and remains in the $\mathrm{Idle}$ state until the timer is triggered. 
The timer deadline is set as
\begin{equation}
  ddl_i(t) = t + \delta^{\mathrm{tx}}_{l_{\max}}+\delta^{\mathrm{tx}}_{\mathrm{ack}} + 2 \delta^{\mathrm{prop}}_{i,m} + \delta^{\mathrm{guard}},
  \label{timer}
\end{equation}
where $\delta^{\mathrm{prop}}_{i,m}$ is the propagation delay estimated from the last successful communication between transmitter $i$ and sink $m$ (continuously updated to reflect location fluctuations), 
and $\delta^{\mathrm{guard}}$ is a guard interval that accounts for propagation uncertainty.
Since no transmission is scheduled in this slot, the feedback is set to $f_i(t)=0$, and transmitter $i$ enters the next slot to make a new decision when the timer expires at $ddl_i(t)$.

\subsubsection{Transmission ($u_i^{\mathrm{exe}}=1$)}
If transmitter $i$ executes transmission at time $t$, it first stays in the $\mathrm{Idle}$ state for a duration of $\tau_i$ and then transmits a data packet of size $l_i$. 
At $t + \tau_i$, it initiates a timer with deadline $ddl_i(t+\tau_i)$ as \eqref{timer}.
After sending the data packet, transmitter $i$ returns to the $\mathrm{Idle}$ state and waits for the corresponding ACK.
Transmitter $i$ may obtain either of the following two feedback outcomes:
\begin{itemize}
  \item \textit{ACK Reception:} if the ACK arrives before $ddl_i(t+\tau_i)$, transmitter $i$ cancels the timer, sets $f_i(t)=1$, and immediately proceeds to the next slot.
  \item \textit{ACK Loss:} if no ACK is received by $ddl_i(t+\tau_i)$, transmitter $i$ sets $f_i(t)=-1$ and enters the next slot when the timer expires.
\end{itemize}
The triggered slot transitions under transmission are depicted in Fig.~\ref{tts}\subref{tts2}--\subref{tts3}.

As illustrated in \eqref{timer}, the timer's deadline is positively correlated with the propagation delay, allowing transmitters closer to the sink node to make decisions more frequently.
This effectively takes advantage of the propagation delay differences between each transmitter and the sink node.
Moreover, the triggered slot paradigm eliminates the limitation of synchronized slot with fixed duration to allow transmitters to freely choose when to transmit data, thereby improving the flexibility of channel utilization.

\begin{figure*}
  \centerline{\includegraphics[width=0.7\textwidth]{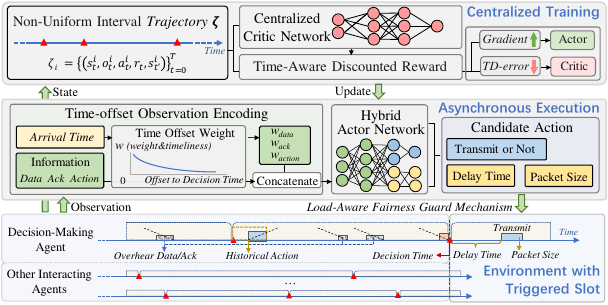}}
  \caption{The flowchart of the proposed time-offset asynchronous MAPPO. Agents execute asynchronously under the triggered slot paradigm and generate candidate actions, while centralized training leverages global states to update the policy.}
  \label{amappo}
\end{figure*}

\subsection{Time-offset Asynchronous MAPPO \label{sec:mappo}}

To fully exploit our proposed triggered slot paradigm and maximize the long-term effective utilization objective $J_U(\pi)$ in \eqref{eq:JU}, we leverage multi-agent reinforcement learning (MARL) to coordinate cooperative decision-making among decentralized transmitters.
However, the introduction of the triggered slot breaks the conventional MARL assumption of synchronous decision-making across agents, creating a fundamental mismatch between protocol design and learning framework.
To accommodate asynchronous decision-making enabled by the triggered slot, we model the multi-node cooperative channel access problem in UANs as a Time-offset Decentralized Partially Observable Markov Decision Process (To-Dec-POMDP), and develop a time-offset asynchronous MAPPO algorithm to solve it.
The flowchart of the proposed time-offset asynchronous MAPPO is illustrated in Fig.~\ref{amappo}.

To-Dec-POMDP can be represented as a tuple $<\mathcal{S}, \mathcal{O}, \mathcal{A}, P, \mathcal{R},  N, \gamma>$,
where $s(t) \in \mathcal{S}$ is the environment state at time $t$, and $o_i(t) \in \mathcal{O}$ is the local observation of agent $i$ at time $t$.
Agent $i$ selects action $a_i(t) \in \mathcal{A}$ based on observation $o_i(t)$.
The environment transitions to the next state $s(t')$ and returns a reward $r=\mathcal{R}(s,a_i)$.

\textbf{Async Hybrid Action Space $\mathcal{A}$}.
All transmitters share the same action structure and select actions asynchronously.
As described in the triggered slot, the action of a transmitter is defined as a tuple consisting of three elements: whether to transmit, the delay before transmission, and the data packet size.
Accordingly, the action space is formulated as:
\begin{equation}
  \mathcal{A} = \{0,1\} \times [0, \tau_{\max}] \times [0, \min \{l_{\max}, q_i(t)\}].
  \label{action}
\end{equation}

At any given time $t^{\mathrm{act}}_i$, agent $i$ selects a hybrid action $a_i = \big(u_i,\, \tau_i,\, l_i\big)$, which is recorded as $\big(t^{\mathrm{act}}_i, a_i\big)$.
Transmitters act asynchronously; yet the long propagation delay couples their actions, and the resulting joint behavior determines the environment state and reward.

\textbf{Time-offset Observation Space $\mathcal{O}$.}
The global environment state $s(t)$ is the concatenation of all agent observations, denoted as $s(t) = [o_1(t), o_2(t), \cdots, o_N(t)]$.
However, due to the long propagation delay and half-duplex operation, each transmitter can only obtain partial and delayed information about the network status, leading to a partial observation.
Hence, we construct the observation $o_i(t)$ of agent $i$ at time $t$ from three event histories in the local channel:
\begin{equation}
  o_i(t)
  =
  \mathcal{H}_i^{\mathrm{act}}
  \ \cup\
  \mathcal{H}_i^{\mathrm{data}}
  \ \cup\
  \mathcal{H}_i^{\mathrm{ack}},
  \label{eq:obs_union}
\end{equation}
where $\mathcal{H}_i^{\mathrm{act}}$, $\mathcal{H}_i^{\mathrm{data}}$, and $\mathcal{H}_i^{\mathrm{ack}}$ are the local histories of agent $i$'s own actions, overheard data, and overheard ACK, respectively.
Such historical events implicitly contain the spatio-temporal topology and connectivity information of the network, which can help intelligent transmitters infer the network status and other transmitters' behavior.
Each history keeps at most $L \times N$ latest events, where $N$ is the number of transmitters and $L$ is a tunable history length controlling the trade-off between observation richness and complexity.

Specifically, the three types of events are recorded as follows:
(1) $\mathcal{H}_i^{\mathrm{act}}$: $(t_i^{\mathrm{act}},\, a_i,\, f_i)$, indicating that transmitter $i$ takes action $a_i$ at time $t_i^{\mathrm{act}}$ and receives feedback $f_i$;
(2) $\mathcal{H}_i^{\mathrm{data}}$: $(t_i^{\mathrm{data}},\, j,\, l)$, indicating that transmitter $i$ receives a data packet sent by transmitter $j\in\mathcal{N}$ with packet size $l$ at time $t_i^{\mathrm{data}}$;
(3) $\mathcal{H}_i^{\mathrm{ack}}$: $(t_i^{\mathrm{ack}},\, j)$, indicating that transmitter $i$ receives an ACK corresponding to a packet from transmitter $j$ at time $t_i^{\mathrm{ack}}$.

\textbf{Time-offset Observation Encoding}. The time-offset mechanism is proposed to provide effective observations for agents at any given time.
Since the time in the observation is represented as the absolute value of the network runtime, it leads to a non-uniform and unbounded observation space, which is not conducive to transmission strategy training.
To address this issue, time offset weight is defined using \textit{linear decay}. 
At any given observation time $t$, each timestamp $t^s$ in the observation is replaced by its corresponding time-offset weights $w(t,t^s)$, computed as
\begin{equation}
  w(t,t^s)
  =
  \max\!\left\{0,\ 1-\frac{|t-t^s|}{L\cdot \tau_{\max}}\right\},
  \label{eq:weight}
\end{equation}
where $t^s$ denotes a timestamp recorded in one of the observation histories, i.e.,$t^\mathrm{act}$ in $\mathcal{H}_i^{\mathrm{act}}$, $t^\mathrm{data}$ in $\mathcal{H}_i^{\mathrm{data}}$, or $t^\mathrm{ack}$ in $\mathcal{H}_i^{\mathrm{ack}}$.
The time offset weight $w(t,t^s) \in [0,1]$ preserves the timeliness of observations while yielding a consistent observation representation for decisions made at arbitrary asynchronous epochs.

\textbf{Time-offset Reward Function $\mathcal{R}$}.
The reward function evaluates the environmental outcome associated with each policy-selected action and provides guidance for subsequent policy updates.
An individual reward is assigned to each agent at each asynchronous triggered slot rather than using a team reward.
The reward is determined by the feedback resulting from the environment interaction.
The reward function of action $a_i = \{u_i,\, \tau_i,\, l_i\}$ at time $t^\mathrm{act}_i$ is defined as:
\begin{equation}
  r_i =
    \begin{cases}
    0, & f_i = 0,\\[4pt]
    \displaystyle \alpha \,\frac{l_i}{l_{\max}} \,\frac{\tau_{\max}}{t - t_i^{\mathrm{act}}}, 
    & f_i = 1,\\[10pt]
    \displaystyle -\,\frac{\tau_{\max}}{t - t_i^{\mathrm{act}}}, 
    & f_i = -1,
    \end{cases}
  \label{reward}
\end{equation}
where $t-t^\mathrm{act}_i$ is the time offset of the last action time relative to the current time, and $\alpha$ is a reward function coefficient to adjust the emphasis of agents on throughput and success rate.

The optimization goal of AT-MAC is to maximize the cumulative reward throughout the network lifetime.
Hence, we define the time-aware discounted return $R_i$ of transmitter $i$ with non-uniform interaction intervals as:
\begin{equation}
  R_i =
  \sum_{t_i^k \leq T}
  \gamma^{t_i^k/\tau_{\max}}
  r_i(t_i^k),
  \label{long}
\end{equation}
where $t_i^k$ is the $k$-th decision epoch of transmitter $i$, $r_i(t_i^k)$ is the corresponding reward, and $T$ is the network lifetime.
The discount factor $\gamma\in(0,1)$ is applied according to the elapsed time from the beginning of the episode, using $\tau_{\max}$ as the normalization scale, thereby accounting for non-uniform interaction intervals.
Finally, the network-level objective is defined by aggregating the individual returns as
\begin{equation}
  R=\sum_{i\in\mathcal{N}} R_i.
  \label{eq:return_net}
\end{equation}

We instantiate the above To-Dec-POMDP with a MAPPO-based centralized training and decentralized execution (CTDE) framework.
Each transmitter $i$ uses a decentralized actor $\pi_{\theta_i}(a_i \mid o_i)$ for local action selection, while a centralized critic leverages the global state $s(t)$ during training.
During asynchronous execution, the actor generates a \emph{candidate} action that is filtered by the load-aware fairness guard mechanism before execution (Section~\ref{sec:fair}).

Compared with standard MAPPO, we incorporate two key adaptations to improve stability and efficiency under the asynchronous triggered slot paradigm: the time-offset observation encoding in~\eqref{eq:weight} and the time-aware discounted return formulation in~\eqref{long}--\eqref{eq:return_net}. 
These adaptations improve the consistency of observation representation and the accuracy of credit assignment across asynchronous decision epochs. 
We refer to the resulting method as \emph{time-offset asynchronous MAPPO}. 
The corresponding training and execution workflow is further adapted to this asynchronous setting, as covered in Section~\ref{sec:workflow}.

\subsection{Load-Aware Fairness Guard Mechanism \label{sec:fair}}

To pursue the long-term load-aware fairness objective $J_F(\pi)$ in \eqref{eq:JF} under decentralized execution, we decouple fairness regulation from the MAPPO optimization objective and incorporate it into environment interaction through a load-aware fairness guard mechanism. This design allows MAPPO to focus on channel-access optimization, while the guard infers network-wide fairness from local overhearing and suppresses transmissions that would violate the prescribed tolerance.

\begin{figure}
  \centerline{\includegraphics[width=0.45\textwidth]{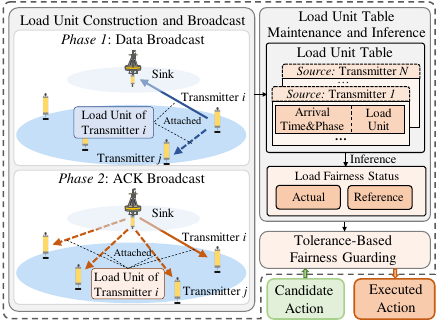}}
  \caption{The flowchart of the load-aware fairness guard mechanism. Each transmitter infers the network-wide load fairness status from local observations, and suppresses transmission decisions from time-offset asynchronous MAPPO that would violate the fairness tolerance.}
  \label{fair}
\end{figure}

The flowchart of the load-aware fairness guard mechanism is illustrated in Fig.~\ref{fair}.
Each transmitter $i$ infers the sent-to-available ratios of other transmitters by leveraging the broadcast nature of the network.
The load-aware fairness guard mechanism facilitates the convergence of these ratios among transmitters, thereby improving the network-wide load-aware fairness.
The complete flow is as follows:

\subsubsection{Load Unit Construction and Broadcast}
At each decision time $t$, transmitter $i$ forms a \emph{load unit}
\begin{equation}
  \mathbf{b}_i(t)=\langle B^{\mathrm{avail}}_i(t),\, B^{\mathrm{deliv}}_i(t)\rangle.
\end{equation}
$B^{\mathrm{avail}}_i(t)$ denotes the available data amount over the observation horizon $h$, and $B^{\mathrm{deliv}}_i(t)$ denotes the successfully delivered data amount over the same horizon, which are calculated according to \eqref{eq:rho}, as:
\begin{equation}
\begin{aligned}
B^{\mathrm{avail}}_i(t) &= q_i(t-h)+\int_{t-h}^{t}\lambda_i(\tau)\,d\tau,\\
B^{\mathrm{deliv}}_i(t) &= \sum_{t_i^k \in [t-h,\,t]} \mathbb{I}\{f_i(t_i^k)=1\}\cdot l_i(t_i^k),
\end{aligned}
\label{eq:AS_def}
\end{equation}
where $q_i(t-h)$ is the queue length at the beginning of the observation horizon, $\lambda_i(\tau)$ is the data arrival rate at time $\tau$, and $t_i^k$ is the $k$-th decision epoch of transmitter $i$.

The load unit $\mathbf{b}_i(t)$ is disseminated through two phases:

(i) Data Broadcast Phase:
During data transmission, transmitter $i$ piggybacks its latest $\mathbf{b}_i(t)$ on the data packet. Since acoustic communication is broadcast, nearby transmitters may overhear this piggybacked load unit even when the data packet is not destined for them;

(ii) ACK Broadcast Phase:
When a data packet is successfully received, the sink relays the parsed load unit in the ACK. 
In the one-hop network, the sink is within the communication range of all transmitters.
Therefore, the ACK phase provides network-wide dissemination, albeit with additional relay latency.

\subsubsection{Load Unit Table Maintenance and Inference}
Each transmitter $i$ maintains a load unit table
\begin{equation}
  \mathcal{T}^{\mathrm{TU}}_{i}=\{\mathcal{B}_{i,1},\mathcal{B}_{i,2},\ldots,\mathcal{B}_{i,N}\},
\end{equation}
where $\mathcal{B}_{i,j}$ stores the load units originated from any other transmitter $j$. 
Each stored record takes the form $(t^{\mathrm{arr}},\, \mathbf{b},\, \mathrm{ph})$,
where $t^{\mathrm{arr}}$ is the reception time at transmitter $i$, $\mathbf{b}$ is the load unit content, and $\mathrm{ph}\in\{\mathrm{D},\mathrm{A}\}$ indicates whether the unit is obtained from the data (D) or ACK (A) broadcast phase.

Due to propagation delay and half-duplex collisions, $\mathcal{T}^{\mathrm{TU}}_{i}$ can be incomplete and delayed. To partially remove phase-dependent offset, transmitter $i$ calibrates each record timestamp to an approximate generation time:
\begin{equation}
  \tilde t =
  \begin{cases}
    t^{\mathrm{arr}} - {2d_{\max}}/{c}, & \mathrm{ph}=\mathrm{A}, \\
    t^{\mathrm{arr}} - {d_{\max}}/{c},  & \mathrm{ph}=\mathrm{D},
  \end{cases}
  \label{eq:time_calib}
\end{equation}
where $d_{\max}$ is the maximum one-hop distance to the sink.

Given the calibrated history in $\mathcal{B}_{i,j}$, transmitter $i$ extrapolates the current load statistics of any other transmitter $j$ at decision time $t$. Let $[\tilde t^{\,\text{last}}_{j},\, B^{\mathrm{avail}}_{j}(t^{\,\text{last}}_{j}),\, B^{\mathrm{deliv}}_{j}(t^{\,\text{last}}_{j})]$ denote the last calibrated record for transmitter $j$. We approximate the current $B^{\mathrm{avail}}_j(t)$ and $B^{\mathrm{deliv}}_j(t)$ as follows:
\begin{equation}
  \begin{aligned}
    \hat B^{\mathrm{avail}}_j(t) &= B^{\mathrm{avail}}_{j}(t^{\,\text{last}}_{j})+ v^{\mathrm{avail}}_{j}(t)\cdot\!\left(t-\tilde t^{\,\text{last}}_{j}\right),\qquad \\
    \hat B^{\mathrm{deliv}}_j(t) &= B^{\mathrm{deliv}}_{j}(t^{\,\text{last}}_{j})+ v^{\mathrm{deliv}}_{j}(t)\cdot\!\left(t-\tilde t^{\,\text{last}}_{j}\right),
  \end{aligned}
  \label{eq:AS_est}
\end{equation}
where $v^{\mathrm{avail}}_{j}(t)$ and $v^{\mathrm{deliv}}_{j}(t)$ are estimated change rates obtained via weighted piecewise linear extrapolation \cite{numerical}. The estimated sent-to-available ratio then follows as
\begin{equation}
  \hat\rho_j(t)=\frac{\hat B^{\mathrm{deliv}}_j(t)}{\hat B^{\mathrm{avail}}_j(t)},
  \quad \forall j \in \mathcal{N} \setminus \{i\}.
\end{equation}
Transmitters with no available data within the fairness horizon are excluded from the load-aware fairness calculation at time $t$.
Finally, transmitter $i$ derives the fairness reference ratio $\hat\rho^{\star}(t)$ based on its locally observed sent-to-available ratio $\rho_i(t)$ and the estimated ratios $\hat\rho_j(t)$ of other transmitters:
\begin{equation}
  \hat\rho^{\star}(t)=\frac{1}{N}\Big(\sum_{j \in \mathcal{N} \setminus \{i\}}\hat\rho_j(t)+\rho_i(t)\Big),
  \label{eq:rho_hat_star}
\end{equation}
which corresponds to the ideal condition that maximizes network-wide load-aware fairness $F(t)$ defined in \eqref{eq:fairness}.

\subsubsection{Tolerance-Based Fairness Guarding}

Given the candidate action $a_i(t)=(u_i(t),\tau_i(t),l_i(t))$ generated by the time-offset asynchronous MAPPO policy, the fairness guard is incorporated into the environment interaction and determines whether the candidate transmission is physically executed.
Specifically, the executed transmission decision $u_i^{\mathrm{exe}}(t)$ is derived from the candidate decision $u_i(t)$ according to
\begin{equation}
  u_i^{\mathrm{exe}}(t)=
  \begin{cases}
    0, &  u_i(t)=1\ \text{and}\ \rho_i(t)>(1+\epsilon)\hat\rho^{\star}(t),\\
    u_i(t), & \text{otherwise},
  \end{cases}
  \label{eq:guard_rule}
\end{equation}
where $\epsilon\ge 0$ is a hyperparameter controlling the tolerance for fairness deviation, with smaller $\epsilon$ imposing stricter fairness requirements but potentially limiting channel utilization.

The fairness-guarding correction is treated as part of the environment interaction. 
The resulting $a_i^{\mathrm{exe}}(t)= (u_i^{\mathrm{exe}}(t),\tau_i(t),l_i(t))$ governs the physical channel access and the corresponding triggered-slot transition, thereby shaping the subsequent feedback and reward.
The original policy-selected action $a_i(t)$ remains unchanged and is used for MAPPO optimization.

\subsection{Workflow of AT-MAC \label{sec:workflow}}

AT-MAC follows the centralized training and decentralized execution framework.
During centralized training, all agents interact asynchronously with a shared simulation environment.
The centralized trainer has access to the global state and the trajectories generated by individual agents to optimize the centralized critic and decentralized actors.
After training, only the actor policy is deployed to each transmitter for decentralized execution, where each agent makes transmission decisions based solely on its local observation.
The workflow of AT-MAC is detailed below.

\subsubsection{Decentralized Action Selection}
Let the decentralized actor of agent $i$ be denoted as $\pi_{\theta_i}(a_i\mid o_i)$ with policy parameters $\theta_i$. 
Given $o_i(t)$, the actor produces a \emph{candidate} action $ a_i(t)=( u_i(t),\tau_i(t),l_i(t))$. 
Since the action contains a discrete transmission indicator and continuous control variables, the discrete component is sampled from its categorical policy, while the continuous components are sampled from Gaussian distributions:
\begin{equation}
    \begin{aligned}
      & u_i(t) \sim \pi_{\theta_i}^{(u)}(\cdot\mid o_i(t)),\\
      & (\tau_i(t),l_i(t)) \sim \mathcal{N}\!(\mu_{\theta_i}\!\big(o_i(t)\big),\,\sigma^2).
    \end{aligned}
  \label{eq:action_select}
\end{equation}
where $\mu_{\theta_i}(\cdot)$ denotes the actor output for the continuous variables, and $\sigma$ denotes the standard deviation used for sampling the continuous actions.
The fairness guard mechanism then maps $a_i(t)$ to $a_i^{\mathrm{exe}}(t)$ according to \eqref{eq:guard_rule}.

\subsubsection{Separate Rollout Buffer}
Each agent interacts with the environment, generating trajectory segments at asynchronous time points $t$:
\begin{equation}
    \zeta_i = \{(o_i(t), a_i(t), r_i(t), o_i(t'))\}_{t=0}^T.
    \label{traj}
\end{equation}
Due to asynchronous decision-making, agents' trajectories are misaligned and may have various lengths.
Therefore, trajectories from multiple agents cannot be aggregated into a shared memory buffer.
To preserve temporally ordered, agent-consistent rollouts, we maintain a separate rollout buffer for each agent, which is essential for correctly computing time-dependent return and advantage estimates (e.g., TD-based or GAE-style) under asynchronous decision epochs.
Each agent's trajectory is aggregated with global state $s(t)$ into its rollout buffer $\mathcal{D}_i$:
\begin{equation}
    \mathcal{D}_i \leftarrow \mathcal{D}_i \cup \{(s(t),o_i(t), a_i(t), r_i(t), s(t'))\}.
    \label{memory}
\end{equation}

\algnewcommand{\LineComment}[1]{\State \(\triangleright\) \textit{#1}} 
 
\begin{algorithm}[t] 
\caption{Workflow of AT-MAC} 
\label{alg:async_mappo} 
\begin{algorithmic}[1] 
\State Initialize actors $\{\pi_{\theta_i}\}_{i=1}^N$ with random weights $\{\theta_i\}_{i=1}^N$ and centralized critic $V_\phi$ with weights $\phi$; 
\State Initialize rollout buffers $\{\mathcal{D}_i\}_{i=1}^N \gets \varnothing$;
\State Set rollout horizon $H$; 
\For{episode $=1$ to $M$} 
      \State Reset the environment and initialize $t=0$;
      \For{each agent $i \in \{1,\dots,N\}$ \textbf{in parallel}}
        \While{$t \leq T_{\max}$} 
            \State Advance time to the decision epoch $t$ of agent $i$; 
            \State Obtain observation $o_i(t)$ and global state $s(t)$; 
            \State Sample action $a_i(t)$ according to \eqref{eq:action_select}; 
            \State Update load unit table $\mathcal{T}^{\mathrm{TU}}_{i}$ using received packets; 
            \State Correct $a_i(t) \to a^{\mathrm{exe}}_i(t)$ via fairness guard \eqref{eq:guard_rule}; 
            \State Execute $a^{\mathrm{exe}}_i(t)$ and obtain $r_i(t)$, $o_i(t')$, and $s(t')$; 
            \State Store $(s(t), o_i(t), a_i(t), r_i(t), s(t'))$ in $\mathcal{D}_i$; 
            \State $o_i(t) \gets o_i(t')$, $s(t) \gets s(t')$, $t \gets t'$; 
            
            \If{$|\mathcal{D}_i| \geq H$} 
              \State Compute advantages $\hat A_i$ and return targets $\hat R_i$;              
              \State Sample mini-batch $\{(s,o,a,r,s')\} \sim \mathcal{D}_i$;  
              \State Update $\phi$ via gradient descent using \eqref{eq:critic_loss};  
              \State Update $\theta_i$ via gradient ascent using \eqref{clip};  
              \State Reset buffer: $\mathcal{D}_i \gets \varnothing$;  
            \EndIf
        \EndWhile 
    \EndFor 
\EndFor 
\end{algorithmic} 
\end{algorithm}

\subsubsection{Centralized Policy Optimization}  
An asynchronous centralized policy optimization strategy is adopted in AT-MAC.  
Specifically, centralized policy training is initiated once each agent's rollout buffer accumulates a trajectory segment of length $H$.
A centralized critic takes the global state $s$ and agent identity $i$ as input to provide an agent-specific value estimate $V_\phi(s,i)$.
The critic is trained by minimizing the time-aware value loss:
\begin{equation}  
  \mathcal{L}_{\mathrm{critic}}(\phi)  
  =\mathbb{E}_{i,t}  
  \left[\left(\hat R_i(t) - V_\phi(s(t),i)\right)^2\right],  
  \label{eq:critic_loss}  
\end{equation}  
where $\hat R_i(t)=\hat A_i(t)+V_\phi(s(t),i)$ is the return target. 
 
The critic's value estimates guide policy updates through Generalized Advantage Estimation (GAE).  
To account for non-uniform decision intervals, the time-aware discount factor is defined as  
$\gamma_t = \gamma^{\Delta t/\tau_{\max}}$, where $\Delta t$ is the time offset between two consecutive decision epochs of agent $i$.  
The temporal-difference residual and advantage estimate are given by
\begin{equation}  
    \delta_i(t) = r_{i}(t)
    +\gamma_t V_\phi(s(t'),i)
    -V_\phi(s(t),i),
\end{equation} 
\begin{equation}  
    \hat A_i(t)=\delta_i(t)+\gamma_t\lambda_{\mathrm{GAE}}\hat A_i(t'),
    \label{advantage}
\end{equation} 
where $\lambda_{\mathrm{GAE}} \in [0,1]$ controls the bias--variance trade-off.  
  
Actors update their policies by maximizing the clipped surrogate objective:  
\begin{equation}  
  \mathcal{L}_{\mathrm{actor}}(\theta_i) =   
  \mathbb{E}_t \Bigg[  
  \begin{aligned}  
      &\min \Big( \rho_t(\theta_i) \hat{A}_i(t), \\  
      &\quad \text{clip} \big( \rho_t(\theta_i), 1-\epsilon_{\text{clip}}, 1+\epsilon_{\text{clip}} \big) \hat{A}_i(t) \Big)  
  \end{aligned}  
  \Bigg]  
\label{clip}  
\end{equation}  
where $\rho_t(\theta_i) =
\pi_{\theta_i}(a_i(t)|o_i(t))/
\pi_{\theta_i^{\mathrm{old}}}(a_i(t)|o_i(t))$
is the importance sampling ratio.  
The clipping range $\epsilon_{\text{clip}}$ constrains policy updates to prevent destructive large steps.  

The trained actor policies are deployed for decentralized execution.
Each agent decides transmission based solely on local partial observations.  

\begin{figure}[tbp]
  \centering
  \captionsetup[subfigure]{font=small} 
  \captionsetup[subfloat]{captionskip=1pt}   
  
  \subfloat[Danjiang Lake, Nanyang, China\label{ft1}]{%
      \includegraphics[width=0.32\linewidth]{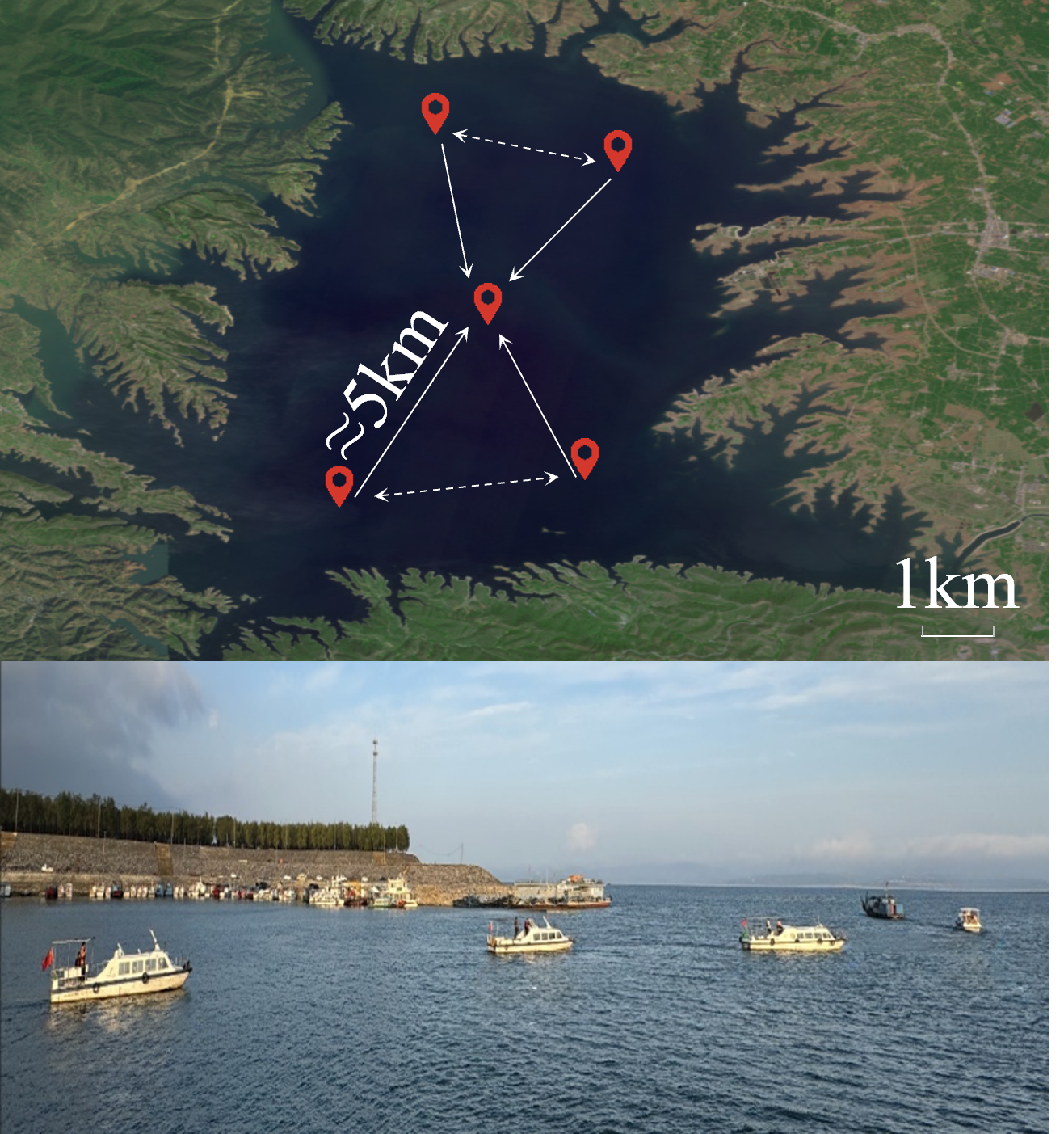}}%
  \hfill
  \subfloat[Jiaozhou Bay, Qingdao, China\label{ft2}]{%
      \includegraphics[width=0.32\linewidth]{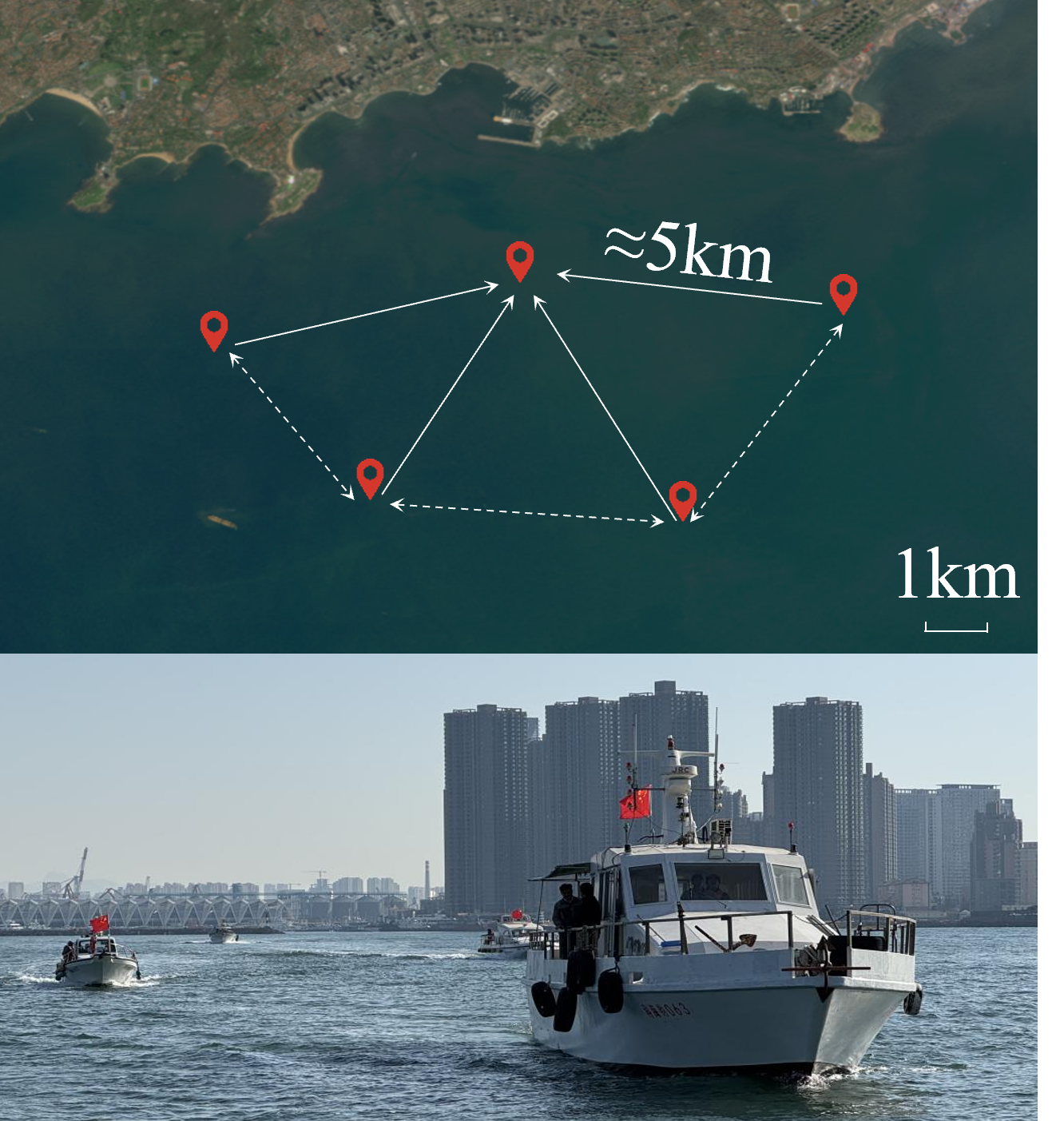}}%
  \hfill
  \subfloat[Songhua Lake, Jilin, China\label{ft3}]{%
      \includegraphics[width=0.32\linewidth]{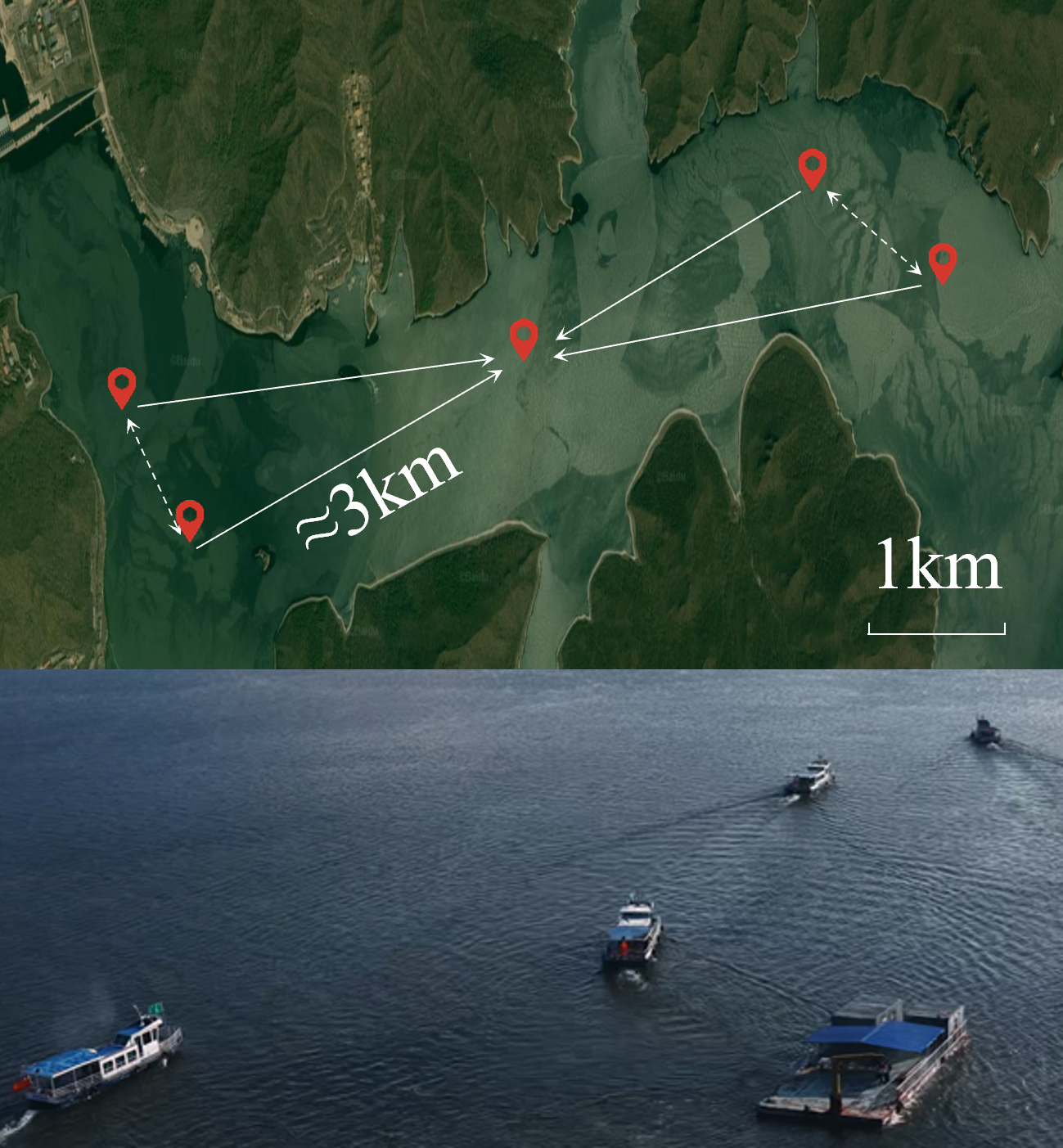}}
  
  \caption{Field trials for collecting real-world data, including: (a) Danjiang Lake trial in October 2024, (b) Jiaozhou Bay trial in March 2025, and (c) Songhua Lake trial in October 2025.}
  \label{ft}
\end{figure}

\section{Field-Reconstructed Validation and Computational Feasibility Analysis}
\label{sec:feasibility}

To evaluate AT-MAC under realistic network configurations, this section constructs simulation scenarios using field-measured network information, including node topology, link connectivity, and hardware communication parameters.
Furthermore, AT-MAC's computational feasibility is evaluated on the embedded platform deployed in field trials.

\subsection{Overview of Field Trials and Hardware Platform}

As illustrated in Fig.~\ref{ft}, we have conducted several multi-node field trials across different lake and coastal environments.

\begin{figure}[t]
  \centering
  \captionsetup[subfigure]{font=small} 
  \captionsetup[subfloat]{captionskip=1pt}   
  
  \subfloat[Underwater node unit\label{node}]{%
      \includegraphics[width=0.4\linewidth]{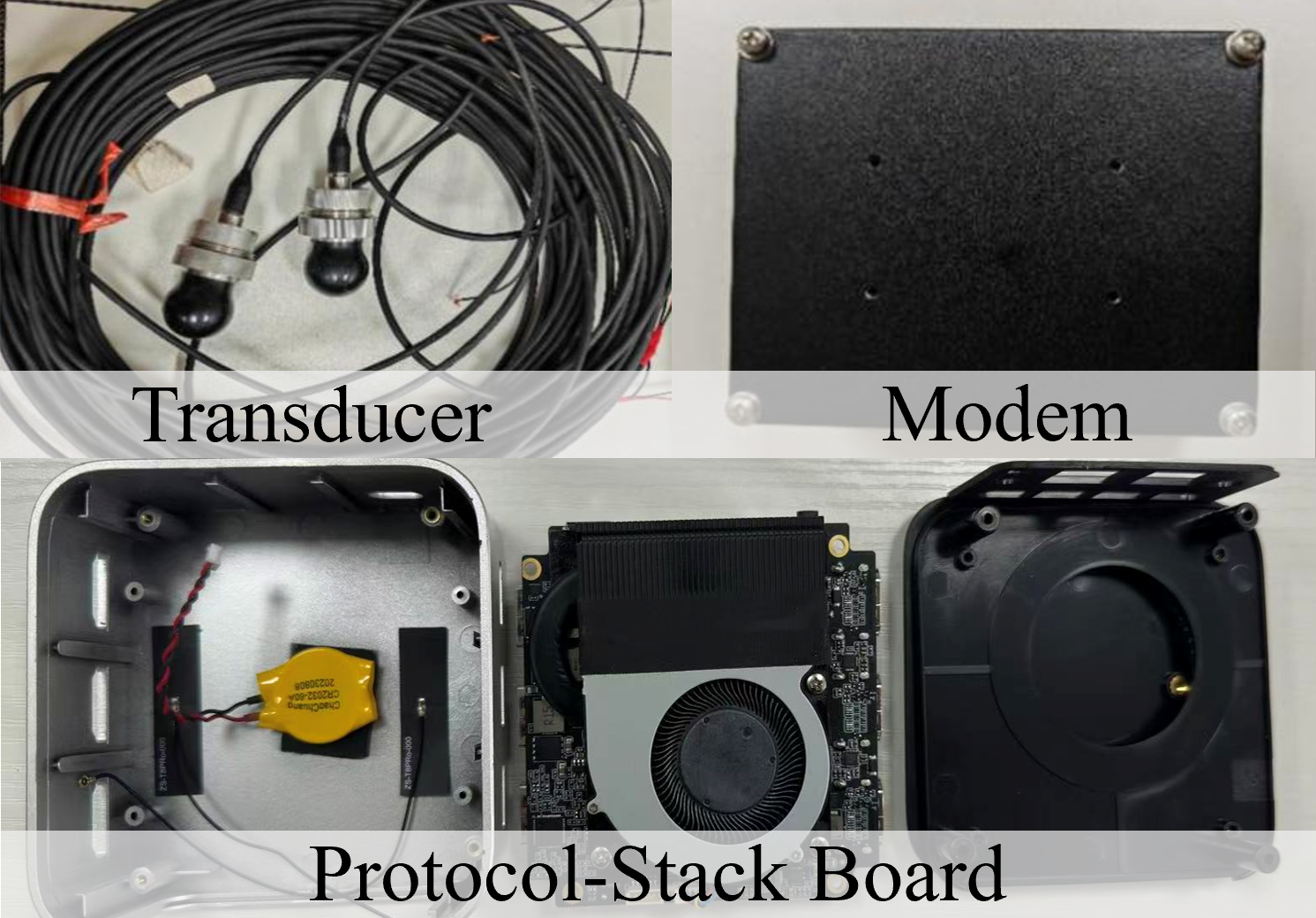}}%
  \hspace{0.5em}
  \subfloat[Shore-based test\label{shore}]{%
      \includegraphics[width=0.4\linewidth]{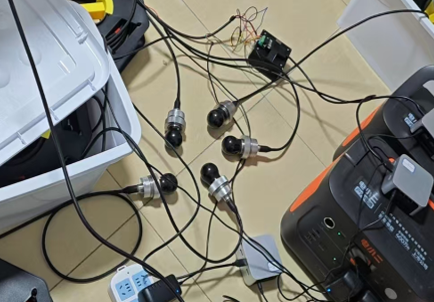}}%

  \caption{Hardware implementation of the underwater node unit and shore-based test scenario.}
  \label{unit}
\end{figure}

The same underwater node unit was utilized across these three trials, comprising: (i) a transducer for underwater electro-acoustic conversion, (ii) an acoustic modem for baseband signal processing, and (iii) an embedded protocol-stack board, as shown in Fig.~\ref{unit}.
The capabilities of the underwater node unit have been validated in multiple field trials, with specific parameters listed in Table~\ref{para}.

\begin{table}[t]
  \centering
  \caption{Communication Parameters}
  \renewcommand{\arraystretch}{1.25}
  \begin{tabular}{ll}
    \hline
    \textbf{Parameter} & \textbf{Value} \\
    \hline
    Effective Transmission Rate & 1000 bps \\
    Maximum Packet Size & 200 Bytes \\
    Preamble Delay & 0.3 s \\
    Stable Maximum Communication Range & 5500 m \\
    \hline
  \end{tabular}
  \label{para}
\end{table}

\subsection{Convergence Validation on Field-Reconstructed Scenarios}

\subsubsection{Scenario Settings}
We reconstruct three simulation scenarios based on field trials conducted at Danjiang Lake, Jiaozhou Bay, and Songhua Lake, each involving a five-node network. 
The nodes were deployed on moored vessels, with their topology and measured connectivity illustrated in Fig.~\ref{ft}.
We employ Aqua-Sim-FG to reconstruct the trial's network environment, including topology, link connectivity, and communication parameters \cite{aqua-sim-fg}.
Aqua-Sim-FG models key underwater acoustic communication characteristics, including half-duplex operation, distance-dependent attenuation, and spatio-temporal packet collisions.
Data generation follows a Poisson process with rate $\lambda^{\mathrm{pkt}}$, which denotes the average arrival rate of maximum-sized packets, while rare stochastic bursts (with probability $1 \times 10^{-4}$) temporarily amplify the arrival rate by 200\% for 1{,}000 s.

\subsubsection{Training Parameters}
The observation history length $L$ is set to $5$, and the reward function coefficient $\alpha$ in \eqref{reward} is set to $2$.
The fairness horizon $h$ in \eqref{eq:rho} is set to $100 \times N$ seconds, and the fairness guard tolerance $\epsilon$ in \eqref{eq:guard_rule} is set to $0.3$.
The chosen parameters are the result of preliminary grid search tuning, which demonstrated robust performance across different scenarios.
Each agent's hybrid actor network comprises two fully connected (FC) layers with 256 hidden units and \textit{Tanh} activation, followed by an output layer composed of three parallel FC layers corresponding to the three elements of the hybrid action.
The critic network has two FC layers with 256 hidden units and a single-unit output layer.
AT-MAC is trained for 10,000 episodes, with the key training parameters summarized in Table~\ref{tp}.

\begin{table}[t]
  \centering
  \caption{Training Parameters}
  \renewcommand{\arraystretch}{1.25}   
  \begin{tabular}{cccc}   
    \hline
    \textbf{Parameters} & \textbf{Values}  &\textbf{Parameters} & \textbf{Values} \\   \hline
    Discount Factor & 0.95 & Actor Learning Rate & $1 \times 10^{-5}$ \\     
    GAE Parameter & 0.95 &  Critic Learning Rate & $1 \times 10^{-4}$ \\   
    Clip Range & 0.2 & Batch Size & 128 \\   
    Entropy Coeff. & 0.01 & Update Horizon & 4096 \\
    Opt. Epochs & 10 & Episode Duration & 10,000 s \\   
    Reward Coeff.  & 2  &  Observation History Length & 5 \\
    Fairness Horizon & $100 \times N$ s & Fairness Guard Tolerance & 0.3 \\
    \hline
  \end{tabular}
  \label{tp}
\end{table}

\subsubsection{Convergence Results}

\begin{figure}[t]
  \centering
  \includegraphics[width=0.7\linewidth]{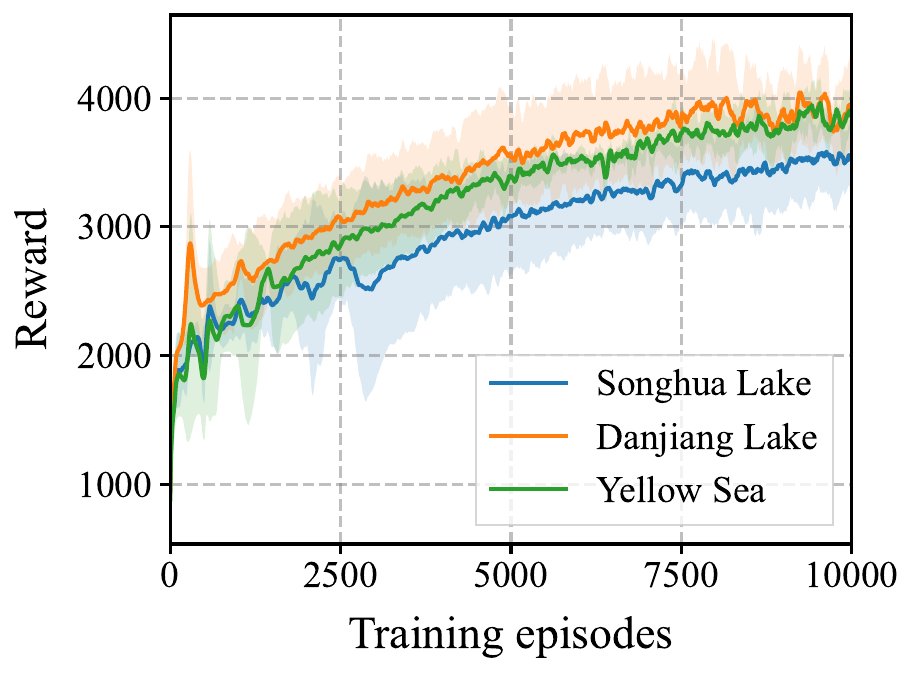}
  \caption{Convergence curves of AT-MAC in the field-reconstructed simulation scenarios. The shaded regions indicate the 95\% confidence intervals.}
  \label{convergence}
\end{figure}

We train AT-MAC 5 times using different random seeds, and report the mean episode reward with the corresponding 95\% confidence interval (shaded region).
Fig.~\ref{convergence} shows the reward convergence of AT-MAC in the reconstructed scenarios.  
The reward increases consistently during training and reaches a clear plateau within 8,000 episodes, while the confidence band progressively narrows and remains stable thereafter.
This pattern indicates that AT-MAC can consistently learn a stable transmission policy across simulation scenarios configured using field-measured topology, connectivity, and modem parameters.

\subsection{Computational Feasibility}

A critical concern for DRL-based protocols is whether the inference overhead is acceptable for resource-constrained underwater nodes.
We evaluate this on the embedded protocol-stack board deployed in the aforementioned field trials.
The board is powered by an N100 processor with a low thermal design power and is responsible for executing the underwater protocol stack and real-time policy inference.

\subsubsection{Model Complexity Analysis}

\begin{figure}[t]
  \centering
  \includegraphics[width=0.7\linewidth]{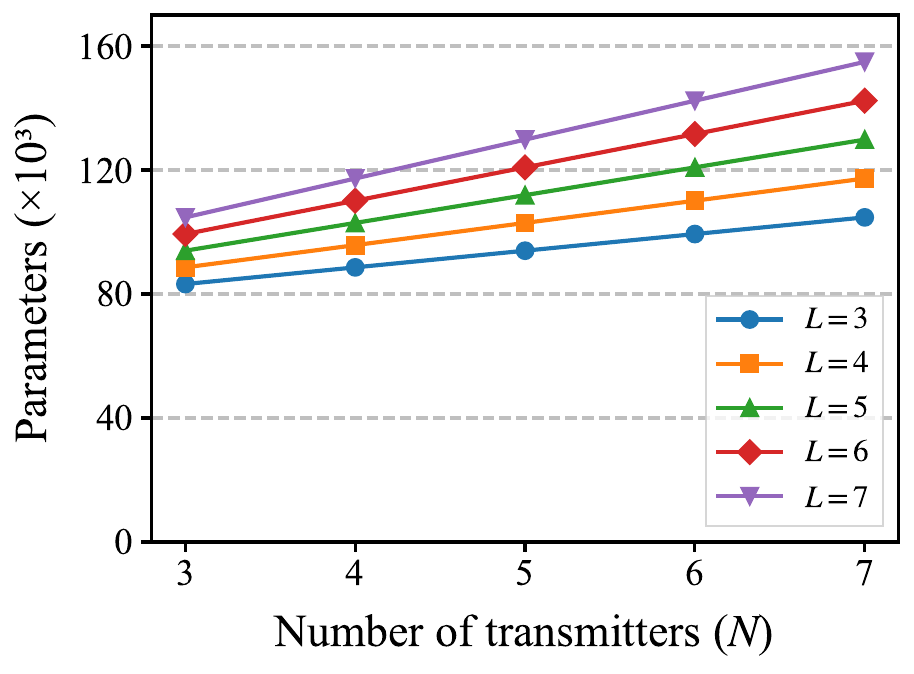}
  \caption{Model complexity of AT-MAC's actor network, measured in the number of parameters across different numbers of transmitters $N$ and observation history lengths $L$.}
  \label{fig:parameters}
  \vspace{-0.8em}
\end{figure}

The Actor network of AT-MAC comprises two shared fully connected (FC) layers with 256 hidden units each (activated by \textit{Tanh}), branching into three parallel FC output heads for hybrid action generation.
Fig.~\ref{fig:parameters} illustrates the parameter count of the Actor network across different network sizes and observation history lengths.
As observed, the parameter count scales linearly with the input dimension, which is determined by the number of transmitters ($N$) and the history window length ($L$). 
Notably, even in the most complex scenario ($N=7, L=7$), the network contains merely $154.88 \times 10^3$ parameters, which requires only $0.31$ M floating-point operations (FLOPs) per inference and occupies approximately $605$ KB of memory.

\subsubsection{On-Board Inference Benchmarking}

We deployed the trained policy network to the embedded protocol-stack board and measured the inference latency over 1,000 independent runs.
The average time required to generate an action is 0.8 $\pm$ 0.1~ms.
Given that acoustic transmission and propagation typically operate on a time scale of seconds, the measured inference latency is several orders of magnitude shorter than the communication time scale.

\section{Performance Evaluation}

This section conducts a comprehensive performance evaluation of AT-MAC using field-reconstructed, dataset-derived, and synthetic scenarios.
Furthermore, we evaluate the effectiveness of the fairness guard through ablation studies and examine the robustness of AT-MAC under clock desynchronization.

The experimental design focuses on four key aspects:
\begin{itemize}
    \item \textbf{Performance and Generalization Profiling:} We evaluate AT-MAC's performance and generalization across varying packet arrival rates in a reconstructed real-world scenario (Section~\ref{sec:perf-profile}).
    \item \textbf{Scalability on Diverse Scenarios:} We evaluate AT-MAC across scenarios, which exhibit different inter-node interference relationships and network scales (Section~\ref{sec:scalability}).
    \item \textbf{Generalization of the Fairness Guard:} We validate the generalization of AT-MAC's fairness guard mechanism under unseen dynamic traffic patterns (Section~\ref{sec:fair_gene}).
    \item \textbf{Ablation and Robustness Studies:} We evaluate the effectiveness of the fairness guard and the robustness of AT-MAC under clock desynchronization (Section~\ref{sec:ablation}).
\end{itemize}

\begin{figure}[t]
  \centering
  \includegraphics[width=1\linewidth]{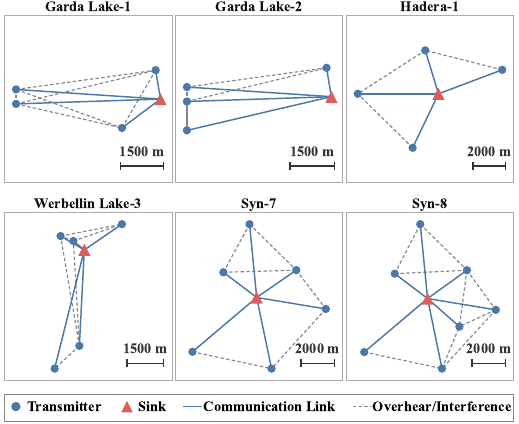}
  \centering 
  \caption{Dataset-derived and synthetic scenarios.}
  \label{fig:more_scenarios}
  \vspace{-1em}
\end{figure}

\subsection{Evaluation Setup}
\subsubsection{Experimental Settings}
AT-MAC is evaluated under three categories of network scenarios, where an $X$-node scenario consists of $X-1$ transmitters and one sink, as summarized below:
\begin{itemize}
  \item \textbf{Field-Reconstructed Scenarios}: This category includes three scenarios from Danjiang Lake (5-node), Jiaozhou Bay (5-node), and Songhua Lake (5-node). The specific topologies and connectivity are illustrated in Fig.~\ref{ft}.
  
  \item \textbf{Dataset-Derived Scenarios}: We utilize the ASUNA dataset~\cite{casari2020asuna}, which provides measured topologies from nearshore trials in Italy, Germany, and Israel. Four topologies are selected: Garda Lake-1 (5-node), Garda Lake-2 (5-node), Hadera-1 (5-node), and Werbellin Lake-3 (6-node), as shown in Fig.~\ref{fig:more_scenarios}. Their relative layouts and connectivity are preserved, with coordinates uniformly scaled to a maximum transmitter-to-sink distance of 5000~m~\cite{comparing}.
  
  \item \textbf{Synthetic Scenarios}: To evaluate scalability beyond the field- and dataset-derived scenarios, we construct two larger synthetic topologies, Syn-7 and Syn-8, as illustrated in Fig.~\ref{fig:more_scenarios}.
\end{itemize}

The communication parameters are consistent with the real field trials, as listed in Table~\ref{para}.
The training parameters for AT-MAC are consistent with those used in the convergence validation, as summarized in Table~\ref{tp}.
All experiments are conducted on a workstation equipped with an \textit{AMD 5995WX} CPU, \textit{128GB} RAM, and an \textit{NVIDIA RTX 4090} GPU.

\begin{figure*}[ht]
  \centering
  \includegraphics[width=0.7\textwidth]{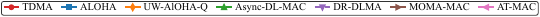} \\
  \vspace{-0.2em} 
  
  \subfloat[Throughput\label{fig:thr_vs_lambda}]{\includegraphics[width=0.35\textwidth]{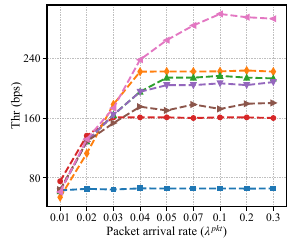}}
  \hspace{1em}
  \subfloat[Success Rate\label{fig:succ_vs_lambda}]{\includegraphics[width=0.35\textwidth]{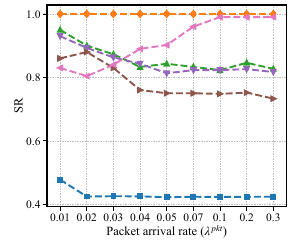}}\\

  \vspace{-0.2em}
  \subfloat[End-to-End Delay\label{fig:delay_vs_lambda}]{\includegraphics[width=0.35\textwidth]{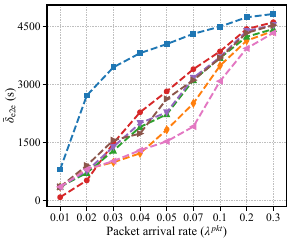}}
  \hspace{1em}
  \subfloat[Fairness\label{fig:fair_vs_lambda}]{\includegraphics[width=0.35\textwidth]{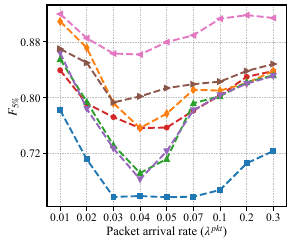}}
  
  \caption{Performance across varying packet arrival rates $\lambda^{\mathrm{pkt}}$ in the reconstructed Danjiang Lake scenario.}
  \label{danjiang}
  \vspace{-1em}
\end{figure*}

\subsubsection{Baseline Methods and Evaluation Metrics}

In our simulation, AT-MAC is compared to six baseline methods under the same simulation scenarios: 
\begin{itemize}
    \item \textbf{TDMA}: A classical protocol that allocates a fixed-length time slot to each transmitter, where each slot is long enough to accommodate a complete data--ACK exchange to avoid collisions.
    
    \item \textbf{ALOHA}: A fundamental random-access protocol where transmitters transmit uncoordinatedly. Unacknowledged packets trigger a random backoff and retransmission, up to a maximum retry limit before being discarded \cite{saloha}.
    
    \item \textbf{UW-ALOHA-Q}: Employs Q-learning to optimize transmission slot selection within a long frame, improving channel temporal reuse and mitigating collisions compared to standard ALOHA \cite{park2019reinforcement}.
    
    \item \textbf{Async-DL-MAC}: Allows transmitters to vary the start timing of data transmissions within a slot. Using a DRL algorithm, it flexibly exploits idle temporal channels caused by the spatial-temporal uncertainty of UANs \cite{exploiting}.
    
    \item \textbf{DR-DLMA}: Divides the channel into micro-slots equal to the packet transmission delay. It utilizes a single-agent DRL algorithm to opportunistically exploit slots left by other transmitters \cite{ye2019deep}.
    
    \item \textbf{MOMA-MAC}: Adopts the short-slot design of DR-DLMA but introduces MAPPO, enabling multiple transmitters to collaboratively learn and exploit idle channel resources to enhance overall network performance \cite{jiang2025underwater}.
\end{itemize}

The evaluation metrics include:
(a) throughput ($\mathrm{Thr}$) in bps, the total successfully delivered data per unit time;
(b) success rate ($\mathrm{SR}$), the ratio of successfully delivered data to total attempted data;
(c) end-to-end delay ($\bar{\delta}_{\text{e2e}}$) in seconds, the byte-weighted mean elapsed time from data generation to successful ACK reception, capturing queuing, transmission, and retransmission overheads;
and (d) load-aware fairness $F_{5\%}$, defined as the 5th percentile of the real-time fairness $F(t)$ in~\eqref{eq:fairness}, which serves as a lower-tail fairness measure.

\subsection{Performance Across Varying Traffic Loads} \label{sec:perf-profile}

This subsection profiles AT-MAC in a reconstructed field-measured scenario (Danjiang Lake) under varying packet arrival rates $\lambda^{\mathrm{pkt}}$.
The policy is trained at $\lambda^{\mathrm{pkt}}=0.1$ and evaluated across a range of $\lambda^{\mathrm{pkt}}$, with each result averaged over five runs with different random seeds, to assess its performance under different traffic loads.

As shown in Fig.~\ref{danjiang}\subref{fig:thr_vs_lambda}, the throughput of all protocols increases with the packet arrival rate $\lambda^{\mathrm{pkt}}$ under light traffic loads, owing to the available channel capacity.
However, as $\lambda^{\mathrm{pkt}}$ increases, the throughput trends diverge significantly, with AT-MAC exhibiting a substantial advantage. Specifically, AT-MAC achieves its peak throughput of 299.5 bps at its training rate ($\lambda^{\mathrm{pkt}}= 0.1$) and maintains superiority across other untrained loads, outperforming the most competitive baselines, UW-ALOHA-Q (222.6 bps) and Async-DL-MAC (216.5 bps), by approximately 34.6\% and 38.3\%, respectively. 
This gain is consistent with the greater scheduling flexibility provided by asynchronous triggered slots, which enables more effective temporal reuse under heterogeneous propagation delays.

\begin{figure*}[ht]
  \centering
  \includegraphics[width=0.7\textwidth]{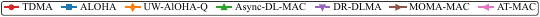} \\
  \vspace{-0.2em} 
  
  \subfloat[Throughput\label{fig:thr_vs_topo}]{\includegraphics[width=0.35\textwidth]{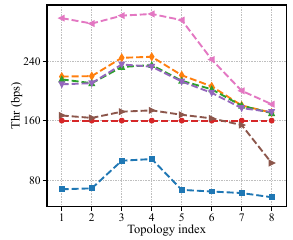}}
  \hspace{1em}
  \subfloat[Success Rate\label{fig:succ_vs_topo}]{\includegraphics[width=0.35\textwidth]{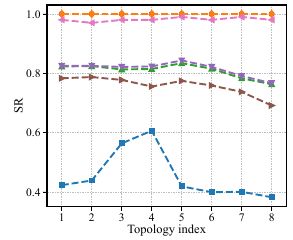}}\\

  \vspace{-0.2em}
  \subfloat[End-to-End Delay\label{fig:delay_vs_topo}]{\includegraphics[width=0.35\textwidth]{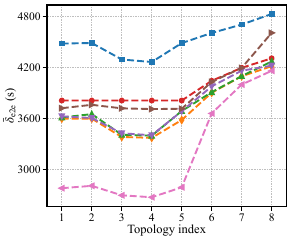}}
  \hspace{1em}
  \subfloat[Fairness\label{fig:fair_vs_topo}]{\includegraphics[width=0.35\textwidth]{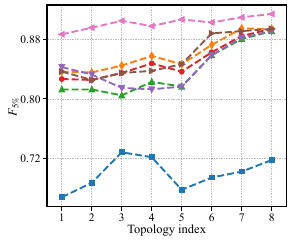}}
  
\caption{Performance across diverse network topologies at $\lambda^{\mathrm{pkt}}=0.1$. Indices 1--8 correspond to: (1) Jiaozhou Bay, 5-node; (2) Songhua Lake, 5-node; (3) Garda Lake-1, 5-node; (4) Garda Lake-2, 5-node; (5) Hadera-1, 5-node; (6) Werbellin Lake-3, 6-node; (7) Syn-7, 7-node; and (8) Syn-8, 8-node.}
  \label{fig:scalability}
  \vspace{-1em}
\end{figure*}

As shown in Fig.~\ref{danjiang}\subref{fig:succ_vs_lambda}, the SR of most baselines deteriorates as $\lambda^{\mathrm{pkt}}$ increases due to intensified channel contention.
In contrast, AT-MAC exhibits a lower SR under light loads (e.g., $83.2\%$ at $\lambda^{\mathrm{pkt}}=0.01$).
This stems from imperfect generalization: trained specifically at $\lambda^{\mathrm{pkt}}=0.1$, the AT-MAC policy yields suboptimal scheduling decisions in low-load states.
However, as $\lambda^{\mathrm{pkt}}$ approaches the training regime, the SR of AT-MAC stabilizes near $100\%$ (e.g., $98.3\%$ at $\lambda^{\mathrm{pkt}}=0.1$).

As shown in Fig.~\ref{danjiang}\subref{fig:delay_vs_lambda}, the end-to-end delay of all protocols increases with $\lambda^{\mathrm{pkt}}$, primarily due to exacerbated queuing delays. 
Across all evaluated load levels, AT-MAC consistently exhibits superior delay performance. 
This advantage is attributed to its superior throughput and high success rate, which alleviate queue congestion and minimize time-consuming retransmissions over long-delay underwater acoustic channels.

As shown in Fig.~\ref{danjiang}\subref{fig:fair_vs_lambda}, the fairness of all protocols shows a non-linear trend with increasing $\lambda^{\mathrm{pkt}}$. 
At $\lambda^{\mathrm{pkt}} = 0.01$, fairness is high across all protocols (e.g., $0.92$ for AT-MAC) due to sufficient channel capacity. 
At medium loads ($\lambda^{\mathrm{pkt}} \in [0.02, 0.07]$), AT-MAC experiences a slight fairness drop (decreasing to $0.86$ at $\lambda^{\mathrm{pkt}} = 0.04$). 
This occurs because its fairness guard mechanism effectively suppresses greedy nodes but cannot actively compensate disadvantaged ones, reflecting imperfect generalization under non-training loads. 
As $\lambda^{\mathrm{pkt}}$ increases further, the fairness of all protocols exhibits an upward trend, since severe congestion uniformly constrains the resource acquisition of all nodes. 
Despite this mid-load fluctuation, AT-MAC consistently outperforms all baselines. 
This confirms that the proposed fairness guard mechanism successfully mitigates disproportionate channel occupation to maintain network-wide fairness.

Overall, the profiling results demonstrate that despite minor generalization limitations under non-training regimes, AT-MAC consistently achieves superior performance across varying traffic loads.

\subsection{Performance Across Diverse Topologies and Network Sizes}\label{sec:scalability}

This subsection evaluates the performance of AT-MAC across a diverse set of network topologies, including field-reconstructed topologies (Jiaozhou Bay and Songhua Lake), public dataset-derived topologies from ASUNA (Garda Lake 1/2, Hadera 1, and Werbellin Lake 3), and two larger synthetic topologies (Syn-7 and Syn-8). The packet arrival rate is fixed at $\lambda^{\mathrm{pkt}}=0.1$.
For each topology, AT-MAC is independently trained and evaluated at $\lambda^{\mathrm{pkt}}=0.1$, with each result averaged over five runs with different random seeds.

As shown in Fig.~\ref{fig:scalability}\subref{fig:thr_vs_topo}, AT-MAC achieves the highest throughput across all evaluated topologies, peaking at 303.27 bps in topology 4 (Garda Lake-2). 
However, its throughput advantage over TDMA narrows in the larger synthetic topologies. 
In Syn-8, AT-MAC achieves 182.04 bps compared with 160.00 bps for TDMA, suggesting that increased interference relationships reduce the opportunities for concurrent transmissions.

Furthermore, most protocols perform better in topologies 3 and 4 (Garda Lake-1 and Garda Lake-2) compared to topologies 1, 2, and 5 (Jiaozhou Bay, Songhua Lake, and Hadera-1), despite all containing 5 nodes. 
The higher throughput in the Garda Lake topologies is consistent with their more heterogeneous transmitter-to-sink propagation delays, which can provide additional opportunities for temporal reuse.

\begin{table*}[t]
\centering
\caption{Packet arrival rates ($\lambda^{\mathrm{pkt}}$) of each transmitter under different traffic load patterns.}
\label{traffic-para}
\setlength{\tabcolsep}{4pt} 
\begin{tabular}{c|cccc|cccc|cccc}
\toprule
 & \multicolumn{4}{c|}{Alternating} 
 & \multicolumn{4}{c|}{Burst} 
 & \multicolumn{4}{c}{Random} \\
Node & Phase 1 & Phase 2 & Phase 3 & Phase 4 & Phase 1 & Phase 2 & Phase 3 & Phase 4 & Phase 1 & Phase 2 & Phase 3 & Phase 4 \\
\midrule
1 & 0.04 & 0.06 & 0.06 & 0   & 0.04 & 0.04  & 0.04  & 0.04  & 0.04 & 0.04  & 0.04  & 0.06 \\
2 & 0.04 & 0.06 & 0   & 0.06 & 0.04 & 0.04  & 0.08 & 0.08 & 0.04 & 0.06 & 0.04  & 0.04  \\
3 & 0.04 & 0   & 0.06 & 0   & 0.04 & 0.04  & 0.04  & 0.04  & 0.04 & 0.04  & 0.04  & 0.06 \\
4 & 0.04 & 0   & 0   & 0.06 & 0.04 & 0.04  & 0.04  & 0.04  & 0.04 & 0   & 0.08 & 0.04  \\

\bottomrule
\end{tabular}
\vspace{-1em}
\end{table*}

\begin{figure*}[ht]
  \centering
  \subfloat[Alternating\label{fair1}]{\includegraphics[width=0.33\textwidth]{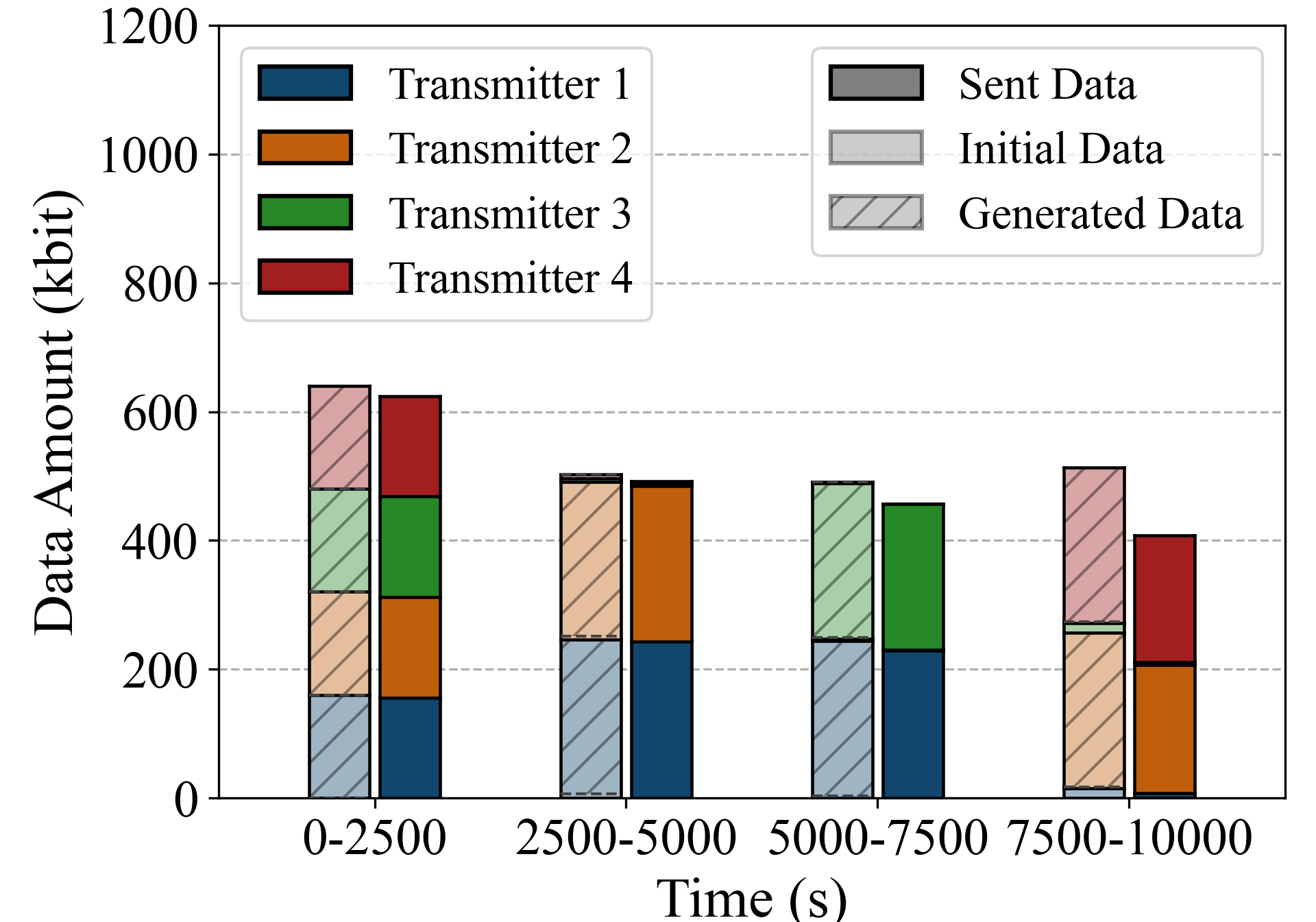}}
  \subfloat[Burst\label{fair2}]{\includegraphics[width=0.33\textwidth]{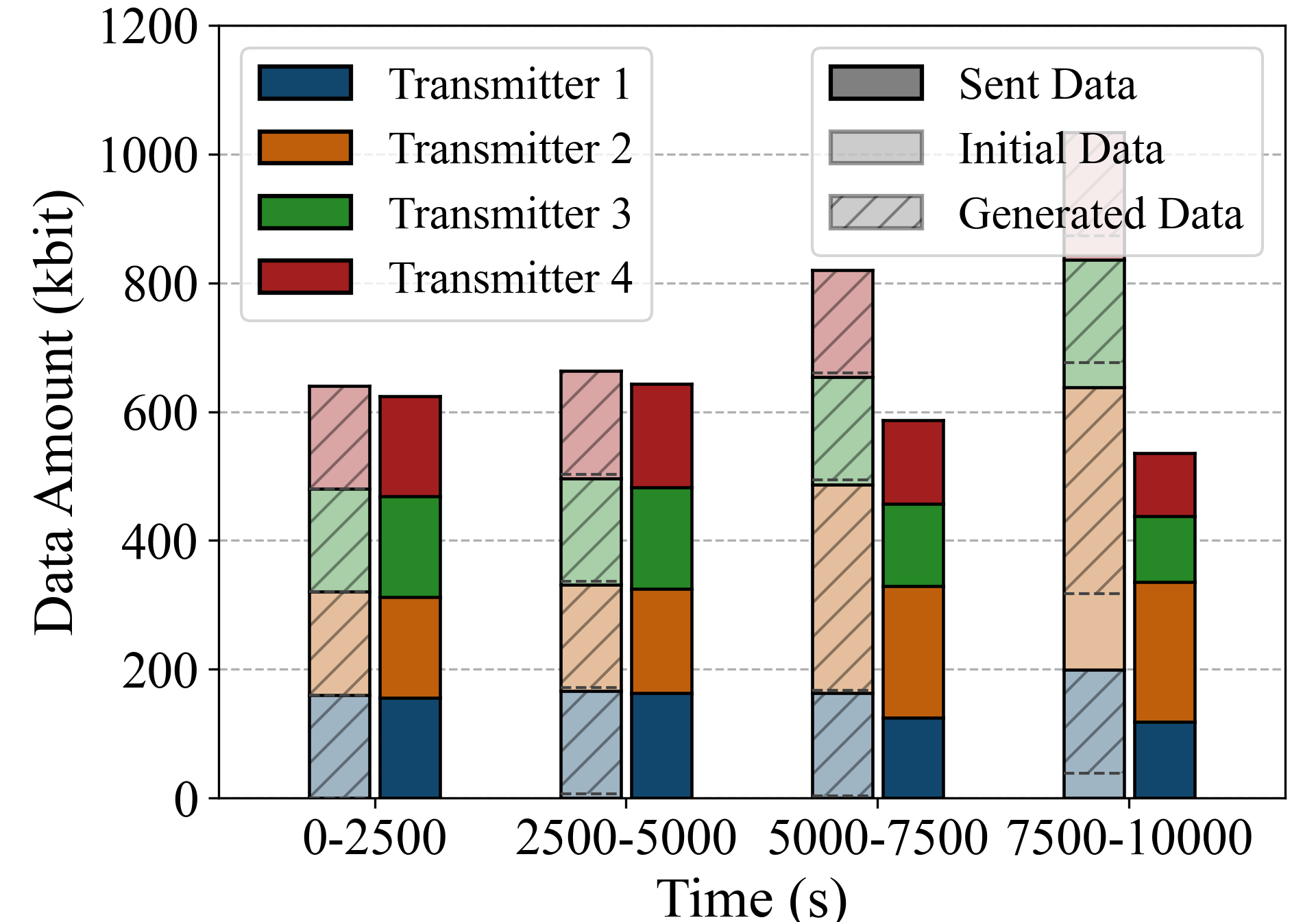}}
  \subfloat[Random\label{fair3}]{\includegraphics[width=0.33\textwidth]{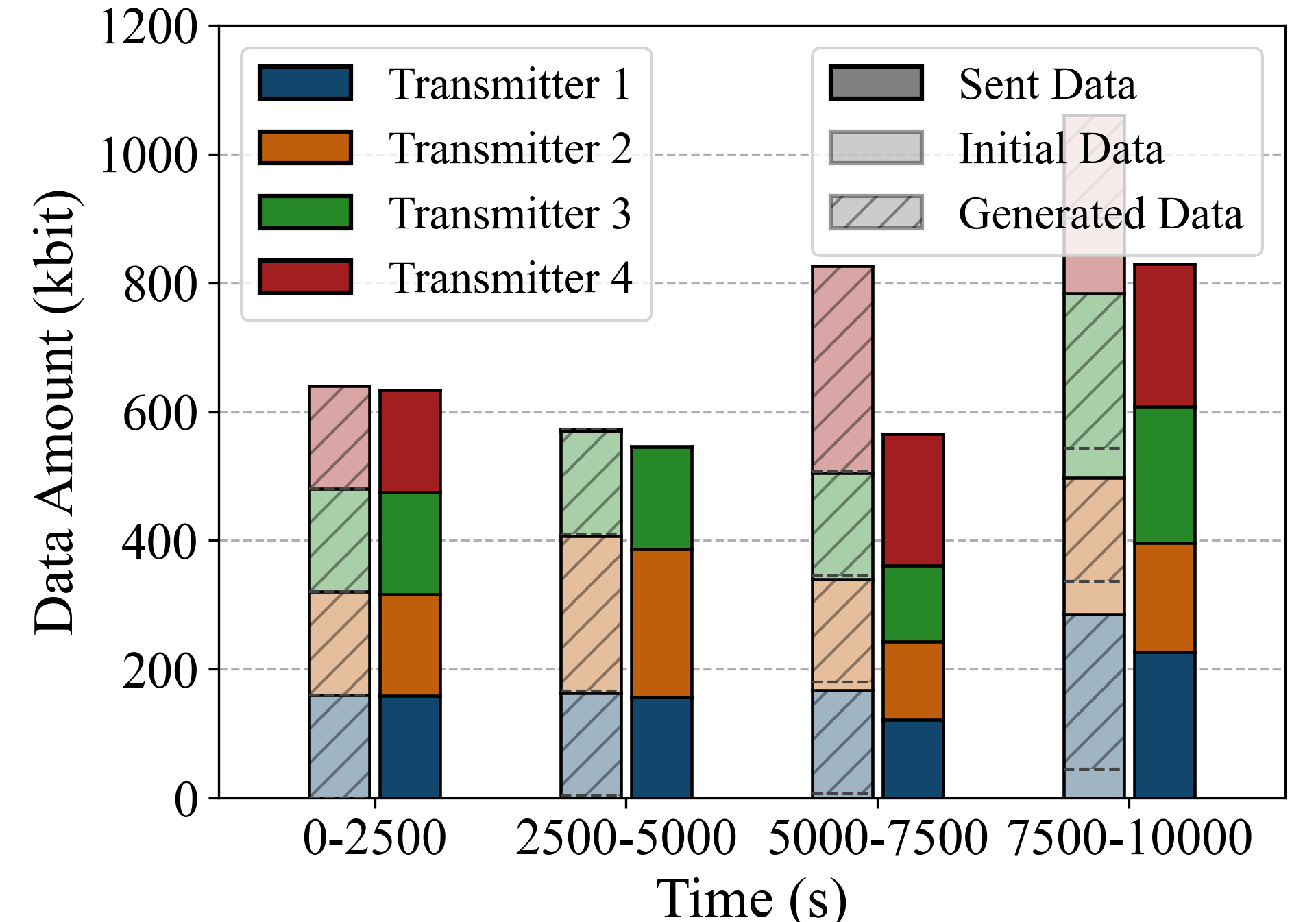}}

  \caption{The generalization of AT-MAC under various traffic load patterns, comparing three metrics in each phase: Initial (backlog at phase start), Generated (new arrivals during the phase), and Sent (successfully delivered data). 
  Fairness is maximized when all transmitters deliver data in equal proportion to their available data (initial plus generated).
  }

  \label{fairness_in_3traffic}
  \vspace{-1em}
\end{figure*}

As shown in Fig.~\ref{fig:scalability}\subref{fig:succ_vs_topo}, AT-MAC maintains a stable SR of approximately 98.0\% across all evaluated topologies, indicating that its asynchronous scheduling remains effective under different interference relationships and network sizes.
As illustrated in Fig.~\ref{fig:scalability}\subref{fig:delay_vs_topo}, the end-to-end delay of all protocols escalates with increasing network scale.
Constrained by the low throughput and success rate caused by more complex interference relationships, all protocols face more severe queue congestion and retransmission overheads in larger networks, leading to significantly increased end-to-end delays.
Nevertheless, AT-MAC consistently achieves the lowest delay, although its delay increases to 4161.40 s in the largest topology (Syn-8), closely approaching TDMA's 4308.67 s.

As shown in Fig.~\ref{fig:scalability}\subref{fig:fair_vs_topo}, the fairness of all protocols generally improves as the network size increases. 
This trend is mainly because severe contention and increased interference in larger networks equally limit the transmission opportunities of all nodes. 
Nevertheless, AT-MAC consistently outperforms the baselines across all evaluated topologies, achieving a high fairness index $F_{5\%}$ of 0.9138 in the largest topology (Syn-8). 
This demonstrates that its fairness guard mechanism effectively maintains fair network-wide channel access, even in complex interference environments.

Overall, AT-MAC maintains a performance advantage across the evaluated topologies. However, this advantage narrows in the larger synthetic scenarios, indicating that increased interference reduces the opportunities for asynchronous channel reuse.

\subsection{Generalization Across Different Traffic Patterns}  \label{sec:fair_gene}

In UANs, the spatial diversity of deployment and the dynamic nature of tasks lead to variable traffic patterns among transmitters.
To evaluate the generalization of AT-MAC under various traffic patterns, we conduct tests in the reconstructed Danjiang Lake scenario, configured with three distinct traffic patterns as follows:
\begin{itemize}
  \item Alternating: Different subsets of transmitters alternate in packet generation, simulating periodic task shifts.

  \item Burst: A transmitter experiences a sharp increase in packet generation rate, emulating a sudden traffic surge.

  \item Random: All transmitters exhibit time-varying packet generation rates, representing an unpredictable condition.
\end{itemize}
The simulation runs for 10,000 seconds, divided into four phases of 2,500 seconds each.
The AT-MAC policy is trained at $\lambda^{\mathrm{pkt}}=0.1$, and then evaluated under the above traffic patterns without further training to assess its generalization to unseen traffic dynamics.
The detailed traffic pattern settings are shown in Table~\ref{traffic-para}.

Fig.~\ref{fairness_in_3traffic} illustrates the generalization of AT-MAC under different traffic patterns.
i) In the alternating pattern, starting from phase 2, different pairs of transmitters alternately generate data.
As shown in Fig.~\ref{fairness_in_3traffic}\subref{fair1}, AT-MAC reallocates the released channel opportunities to the active transmitters, allowing the delivered traffic to adapt to the changes in available data across phases.
ii) In the burst pattern, transmitter 2 experiences a sharp increase in data generation during phases 2--4.
As shown in Fig.~\ref{fairness_in_3traffic}\subref{fair2}, AT-MAC correspondingly allocates more transmission opportunities to transmitter 2, making its delivered-data share increase with its available-data share.
iii) In the random pattern, all transmitters exhibit varying $\lambda^{\mathrm{pkt}}$ during phases 2--4.
Fig.~\ref{fairness_in_3traffic}\subref{fair3} shows that AT-MAC dynamically adjusts the transmission opportunities according to load variations, such that the distribution of successfully delivered data follows the changes in available traffic among transmitters.

Overall, these results demonstrate that AT-MAC adapts its transmission allocation to unseen traffic patterns while preserving load-aware fairness, although some throughput degradation is observed relative to the training condition.

\subsection{Ablation Study}  \label{sec:ablation}

\subsubsection{Effectiveness of the Load-Aware Fairness Guard Mechanism}
To isolate the specific contribution of the load-aware fairness guard mechanism, we conduct an ablation study by evaluating the standard AT-MAC against a variant with this mechanism disabled.
We run both variants in the Danjiang Lake scenario under $\lambda^{\mathrm{pkt}}=0.05$, while the AT-MAC policy is trained at $\lambda^{\mathrm{pkt}}=0.1$.

\begin{figure}[t]
  \centering
  \subfloat[\label{ci}]{\includegraphics[width=0.24\textwidth]{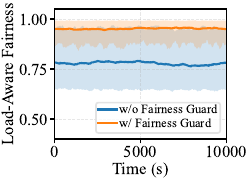}}
  \subfloat[\label{s}]{\includegraphics[width=0.24\textwidth]{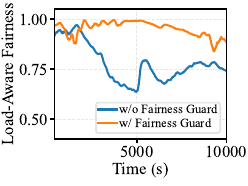}}
  \centering 
  \caption{Fairness over time in the Danjiang Lake scenario with and without the load-aware fairness guard mechanism. (a) The mean fairness across 100 runs with 95\% confidence intervals; (b) Specific fairness curves.}
  \label{fig:ab}
\end{figure}

Fig.~\ref{fig:ab}\subref{ci} shows the mean fairness across 100 independent runs, along with the corresponding 95\% confidence intervals.
With the guard enabled, mean fairness increases from 0.78 to 0.94 and exhibits a substantially narrower 95\% confidence interval. 
Individual runs in Fig.~\ref{fig:ab}\subref{s} further show that the guard suppresses severe fairness degradation: without the guard, $F(t)$ drops to 0.64, while the guarded variant remains comparatively stable. 
These results demonstrate that load-aware transmission correction improves both fairness and its robustness across runs.

\subsubsection{Sensitivity to the Fairness Guard Tolerance}

\begin{figure}[t]
  \centerline{\includegraphics[width=0.65\linewidth]{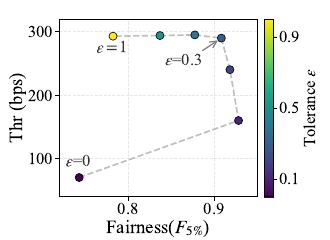}}
  \caption{The sensitivity of AT-MAC's performance to the fairness guard tolerance $\epsilon$.}
  \label{sensitivity}
\end{figure}

To evaluate the impact of the fairness guard tolerance $\epsilon$, we train AT-MAC using various $\epsilon$ values in the Danjiang Lake scenario under $\lambda^{\mathrm{pkt}}=0.1$. 

Fig.~\ref{sensitivity} illustrates the sensitivity of AT-MAC's performance to $\epsilon$.
At $\epsilon=0$, the strict zero tolerance blocks almost all transmission attempts. 
Consequently, the RL agent lacks valid training samples, which prevents the policy from converging and leads to poor throughput and fairness. 
As $\epsilon$ increases, network performance steadily improves. Specifically, $\epsilon=0.3$ provides the best observed trade-off among the evaluated values, achieving the best balance between high throughput and network fairness. 
However, increasing $\epsilon$ beyond 0.3 brings little benefit: the throughput barely grows, while the fairness drops significantly. 
This is because a larger tolerance fails to suppress greedy nodes, rendering the fairness guard ineffective.

\begin{figure}[t]
  \centerline{\includegraphics[width=0.7\linewidth]{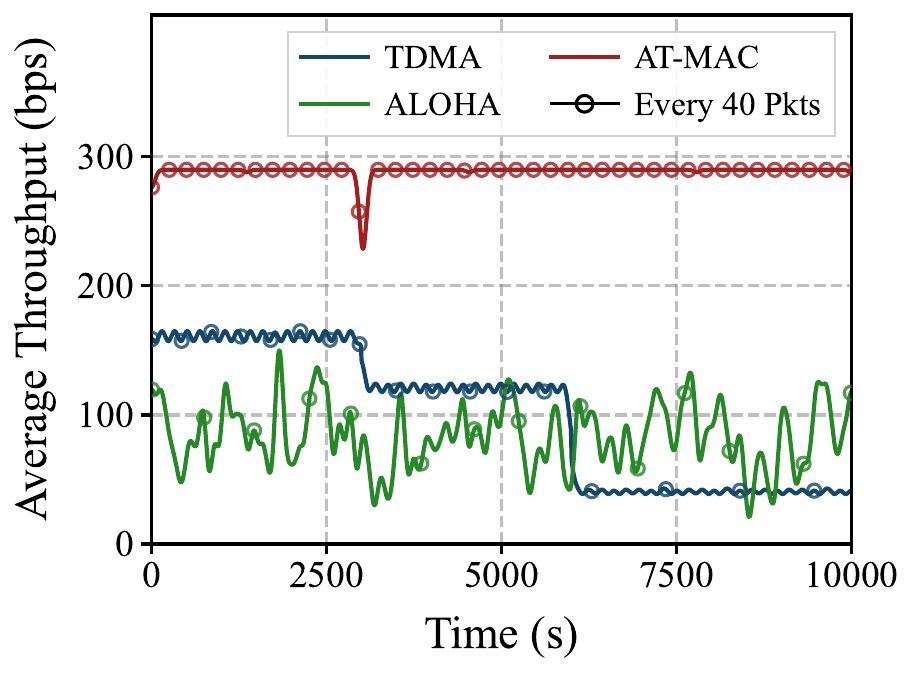}}
  \caption{Average throughput (smoothed by a 40-second sliding window) of each protocol under clock desynchronization.}
  \label{clock}
\end{figure}

\subsubsection{Robustness under Clock Desynchronization}

In underwater acoustic networks, clock desynchronization can disrupt slot alignment and degrade the performance of synchronization-dependent MAC protocols \cite{clock}.
This subsection evaluates the robustness of AT-MAC under clock desynchronization in the Danjiang Lake scenario.
We set up two scenarios of clock desynchronization:
\begin{itemize}
  \item Clock Drift: The node clock drifts at a constant rate, simulating a gradual desynchronization over time. 
  Transmitter 1's clock is behind by 0.001 seconds every second.
  \item Clock Jump: A node undergoes a fault and restarts, causing its clock to jump.
  Transmitter 4's clock jumps forward by 4 seconds at 3000~s.
\end{itemize}

Fig.~\ref{clock} illustrates the average throughput under clock desynchronization. 
TDMA suffers significant performance degradation, dropping to 120 bps at 3000 s due to a sudden clock jump that causes Transmitter 4's packets to collide with Transmitter 3's ACKs, and plummeting to 40 bps at 6000 s as cumulative drift leads to collisions between Transmitters 1 and 2.
Conversely, ALOHA remains largely unaffected (fluctuating between 20-150 bps), as its random access mechanism does not systematically increase collision probability due to unaligned time slots.
AT-MAC maintains a high and stable throughput of 290 bps. 
By utilizing independent timelines, AT-MAC does not rely on aligned slots and therefore remains robust to the resulting slot misalignment.
Although the sudden clock jump at 3000~s temporarily makes part of the observation history temporally inconsistent and causes a brief performance drop, AT-MAC recovers as the observation window is gradually replenished with newly collected observations.

\section{Conclusion}

This paper investigates a transition from synchronized slots to asynchronous scheduling in DRL-based underwater MAC, allowing MAC protocols to better accommodate the inherent asynchrony of underwater acoustic channels.
To realize this transition, we propose an asynchronous triggered MAC, denoted as AT-MAC.
AT-MAC integrates a triggered slot paradigm, an asynchronous time-offset MAPPO algorithm, and a low-overhead load-aware fairness guard mechanism to achieve efficient and fair cooperative channel access using local information.
Field-reconstructed simulations evaluate AT-MAC under network configurations derived from real deployments, while on-board inference benchmarking demonstrates its computational feasibility on the embedded protocol-stack platform. 
Extensive simulations show consistent performance gains across the evaluated traffic loads, topologies, and traffic patterns. 
Ablation experiments validate the fairness guard, while clock-desynchronization tests demonstrate the robustness of AT-MAC.

Future work will explore broader applications of asynchronous event-triggered mechanisms, with particular emphasis on cross-layer designs integrating routing and MAC protocols in multi-hop UANs, thereby enhancing overall performance in more complex network environments.

\section*{Acknowledgments}
This work was supported in part by the National Natural Science Foundation of China under Grant 62471201 and Grant 62501250; in part by the Postdoctoral Science Foundation of China under Grant 2025M771509; in part by the Postdoctoral Fellowship Program of CPSF under Grant Number GZC20250178; in part by the Fundamental Research Funds for the Central Universities under Grant Number 45124031D041.

\bibliographystyle{IEEEtran}
\bibliography{ref}

@article{DOTS,
  title={DOTS: A propagation delay-aware opportunistic MAC protocol for mobile underwater networks},
  author={Noh, Youngtae and Lee, Uichin and Han, Seongwon and Wang, Paul and Torres, Dustin and Kim, Jinwhan and Gerla, Mario},
  journal={IEEE Transactions on Mobile Computing},
  volume={13},
  number={4},
  pages={766--782},
  year={2014},
  publisher={IEEE}
}

@article{jiang2025underwater,
  title={Underwater Acoustic MAC Protocol for Multi-Objective Optimization Based on Multi-Agent Reinforcement Learning},
  author={Jiang, Jinfang and Dong, Yiling and Han, Guangjie and Su, Gang},
  journal={Drones},
  volume={9},
  number={2},
  pages={123},
  year={2025},
  publisher={MDPI}
}

@article{casari2020asuna,
  title={ASUNA: A topology data set for underwater network emulation},
  author={Casari, Paolo and Campagnaro, Filippo and Dubrovinskaya, Elizaveta and Francescon, Roberto and Dagan, Amir and Dahan, Shlomo and Zorzi, Michele and Diamant, Roee},
  journal={IEEE Journal of Oceanic Engineering},
  volume={46},
  number={1},
  pages={307--318},
  year={2020},
  publisher={IEEE}
}

@article{comparing,
  title={Comparing underwater MAC protocols in real sea experiments},
  author={Pu, Lina and Luo, Yu and Mo, Haining and Le, Son and Peng, Zheng and Cui, Jun-Hong and Jiang, Zaihan},
  journal={Computer Communications},
  volume={56},
  pages={47--59},
  year={2015},
  publisher={Elsevier}
}

@article{2013Comparing,
  title={Comparing the SUNSET and DESERT frameworks for in field experiments in underwater acoustic networks},
  author={ Petroccia, Roberto  and  Spaccini, Daniele },
  journal={IEEE},
  year={2013},
}

@book{numerical,
  title={Numerical Analysis},
  author={Burden, Richard L. and Faires, J. Douglas},
  edition={9th},
  year={2010},
  publisher={Brooks/Cole, Cengage Learning},
  address={Boston, MA},
  isbn={9780538733519}
}

@inproceedings{2006Time,
  title={Time Synchronization for High Latency Acoustic Networks},
  author={ Syed, Affan A  and  Heidemann, John },
  booktitle={INFOCOM 2006. 25th IEEE International Conference on Computer Communications, Joint Conference of the IEEE Computer and Communications Societies, 23-29 April 2006, Barcelona, Catalunya, Spain},
  year={2006},
}

@Article{PTA-Sync,
AUTHOR = {Cho, A-Ra and Choi, Youngchol},
TITLE = {PTA-Sync: Packet-Train-Aided Time Synchronization for Underwater Acoustic Applications},
JOURNAL = {Applied Sciences},
VOLUME = {13},
YEAR = {2023},
NUMBER = {2},
ARTICLE-NUMBER = {978},
ISSN = {2076-3417},
DOI = {10.3390/app13020978}
}

@ARTICLE{delayaware,
  author={Huang, Jiajie and Ye, Xiaowen and Wang, Yizhe and Fu, Liqun},
  journal={IEEE Internet of Things Journal}, 
  title={Leveraging Propagation Delays: A Delay-Aware Multi-Agent Reinforcement Learning MAC Protocol for Underwater Acoustic Networks}, 
  year={2025},
  volume={},
  number={},
  pages={1-1},
  doi={10.1109/JIOT.2025.3595133}}

@INPROCEEDINGS{deep2,
  author={Du, Haiyang and Wang, Xiaomei and Sun, Weikai and Zhang, Jiasen},
  booktitle={2024 6th International Conference on Communications, Information System and Computer Engineering (CISCE)}, 
  title={An Adaptive MAC Protocol for Underwater Acoustic Networks Based on Deep Reinforcement Learning}, 
  year={2024},
  volume={},
  number={},
  pages={1271-1275},
  doi={10.1109/CISCE62493.2024.10652998}}

@ARTICLE{subslot,
  author={Xie, Weiliang and Shen, Xiaohong and Sun, Lin and Wang, Chao and Yan, Yongsheng and Wang, Haiyan},
  journal={IEEE Internet of Things Journal}, 
  title={Dynamic Optimization of Slot Management MAC Protocol for Large-Scale IoUT Based on POMDP}, 
  year={2025},
  volume={12},
  number={15},
  pages={29784-29796},
  doi={10.1109/JIOT.2025.3568929}}

@inproceedings{saloha,
  title={Aloha-based MAC protocols with collision avoidance for underwater acoustic networks},
  author={Chirdchoo, Nitthita and Soh, W-S and Chua, Kee Chaing},
  booktitle={IEEE INFOCOM 2007-26th IEEE International Conference on Computer Communications},
  pages={2271--2275},
  year={2007},
  organization={IEEE}
}

@article{TDA,
  title={TDA-MAC: TDMA without clock synchronization in underwater acoustic networks},
  author={Morozs, Nils and Mitchell, Paul and Zakharov, Yuriy V},
  journal={IEEE Access},
  volume={6},
  pages={1091--1108},
  year={2017},
  publisher={IEEE}
}

@article{survey,
  title={A survey on MAC protocols for underwater wireless sensor networks},
  author={Chen, Keyu and Ma, Maode and Cheng, En and Yuan, Fei and Su, Wei},
  journal={IEEE Communications Surveys \& Tutorials},
  volume={16},
  number={3},
  pages={1433--1447},
  year={2014},
  publisher={IEEE}
}

@inproceedings{stump,
  title={STUMP: Exploiting position diversity in the staggered TDMA underwater MAC protocol},
  author={Kredo II, Kurtis and Djukic, Petar and Mohapatra, Prasant},
  booktitle={IEEE INFOCOM 2009},
  pages={2961--2965},
  year={2009},
  organization={IEEE}
}

@inproceedings{lt,
  title={LT-MAC: A location-based TDMA MAC protocol for small-scale underwater sensor networks},
  author={Mao, Jia and Chen, Shumin and Liu, Yangxi and Yu, Juntao and Xu, Yuanxin},
  booktitle={2015 IEEE International Conference on Cyber Technology in Automation, Control, and Intelligent Systems (CYBER)},
  pages={1275--1280},
  year={2015},
  organization={IEEE}
}

@article{exploiting,
  title={Exploiting propagation delay in underwater acoustic communication networks via deep reinforcement learning},
  author={Geng, Xuan and Zheng, Yahong Rosa},
  journal={IEEE Transactions on Neural Networks and Learning Systems},
  volume={34},
  number={12},
  pages={10626--10637},
  year={2022},
  publisher={IEEE}
}

@article{UIOT,
  title={Underwater Internet of Things in smart ocean: System architecture and open issues},
  author={Qiu, Tie and Zhao, Zhao and Zhang, Tong and Chen, Chen and Chen, CL Philip},
  journal={IEEE transactions on industrial informatics},
  volume={16},
  number={7},
  pages={4297--4307},
  year={2019},
  publisher={IEEE}
}

@article{IOUT,
  title={Internet of underwater things and big marine data analytics—a comprehensive survey},
  author={Jahanbakht, Mohammad and Xiang, Wei and Hanzo, Lajos and Azghadi, Mostafa Rahimi},
  journal={IEEE Communications Surveys \& Tutorials},
  volume={23},
  number={2},
  pages={904--956},
  year={2021},
  publisher={IEEE}
}

@article{ACH,
  title={Achieving Fair-Effective Communications and Robustness in Underwater Acoustic Sensor Networks: A Semi-Cooperative Approach},
  author={Gou, Yu and Zhang, Tong and Liu, Jun and Yang, Tingting and Song, Shanshan and Cui, Jun-Hong},
  journal={IEEE Transactions on Mobile Computing},
  year={2023},
  publisher={IEEE}
}

@article{MCMAC,
  title={An efficient medium access control scheme based on MC-CDMA for mobile underwater acoustic networks},
  author={Guo, Jiani and Song, Shanshan and Liu, Jun and Wan, Lei and Zhao, Yan and Han, Guangjie},
  journal={IEEE Network},
  volume={36},
  number={3},
  pages={167--173},
  year={2022},
  publisher={IEEE}
}

@Article{PFAloha,
AUTHOR = {Alhassan, Ibrahim B. and Mitchell, Paul D.},
TITLE = {Packet Flow Based Reinforcement Learning MAC Protocol for Underwater Acoustic Sensor Networks},
JOURNAL = {Sensors},
VOLUME = {21},
YEAR = {2021},
NUMBER = {7},
ARTICLE-NUMBER = {2284},
PubMedID = {33805233},
ISSN = {1424-8220},
}

@article{DR-ALOHA-Q,
author = {Tomovic, Slavica},
year = {2023},
month = {05},
pages = {4474},
title = {DR-ALOHA-Q: A Q-Learning-Based Adaptive MAC Protocol for Underwater Acoustic Sensor Networks},
volume = {23},
journal = {Sensors},
}

@article{shi2025delay,
  title={Delay-Fluctuation-Resistant Underwater Acoustic Network Access Method Based on Deep Reinforcement Learning},
  author={Shi, Jinli and Tian, Kun and Zhang, Jun},
  journal={Sensors},
  volume={25},
  number={21},
  pages={6673},
  year={2025},
  publisher={MDPI}
}

@ARTICLE{ye2019deep,
  author={Ye, Xiaowen and Yu, Yiding and Fu, Liqun},
  journal={IEEE Transactions on Mobile Computing}, 
  title={Deep Reinforcement Learning Based MAC Protocol for Underwater Acoustic Networks}, 
  year={2022},
  volume={21},
  number={5},
  pages={1625-1638},}

@article{state,
  title={State-of-the-art medium access control (MAC) protocols for underwater acoustic networks: A survey based on a MAC reference model},
  author={Jiang, Shengming},
  journal={IEEE communications surveys \& tutorials},
  volume={20},
  number={1},
  pages={96--131},
  year={2017},
  publisher={IEEE}
}

@article{park2019reinforcement,
  title={Reinforcement learning based MAC protocol (UW-ALOHA-Q) for underwater acoustic sensor networks},
  author={Park, Sung Hyun and Mitchell, Paul Daniel and Grace, David},
  journal={IEEE access},
  volume={7},
  pages={165531--165542},
  year={2019},
  publisher={IEEE}
}

@article{zhang2019load,
  title={A load-based hybrid MAC protocol for underwater wireless sensor networks},
  author={Zhang, Ziwei and Shi, Wei and Niu, Qiuna and Guo, Ying and Wang, Jingjing and Luo, Hanjiang},
  journal={IEEE Access},
  volume={7},
  pages={104542--104552},
  year={2019},
  publisher={IEEE}
}

@article{su2021traffic,
  title={A traffic load-aware OFDMA-based MAC protocol for distributed underwater acoustic sensor networks},
  author={Su, Yishan and Liu, Xuan and Han, Guangyao and Fu, Xiaomei},
  journal={IEEE Transactions on Vehicular Technology},
  volume={70},
  number={10},
  pages={10501--10513},
  year={2021},
  publisher={IEEE}
}

@inproceedings{hsu2009st,
  title={ST-MAC: Spatial-temporal MAC scheduling for underwater sensor networks},
  author={Hsu, C-C and Lai, K-F and Chou, C-F and Lin, KC-J},
  booktitle={IEEE INFOCOM 2009},
  pages={1827--1835},
  year={2009},
  organization={IEEE}
}

@article{park2020reinforcement,
  title={Reinforcement learning based MAC protocol (UW-ALOHA-QM) for mobile underwater acoustic sensor networks},
  author={Park, Sung Hyun and Mitchell, Paul Daniel and Grace, David},
  journal={IEEE Access},
  volume={9},
  pages={5906--5919},
  year={2020},
  publisher={IEEE}
}

@article{gussen2016survey,
  title={A survey of underwater wireless communication technologies},
  author={Gussen, Camila MG and Diniz, Paulo SR and Campos, Marcello LR and Martins, Wallace A and Costa, Felipe M and Gois, Jonathan N},
  journal={J. Commun. Inf. Sys},
  volume={31},
  number={1},
  pages={242--255},
  year={2016}
}

@article{sozer2000underwater,
  title={Underwater acoustic networks},
  author={Sozer, Ethem M and Stojanovic, Milica and Proakis, John G},
  journal={IEEE journal of oceanic engineering},
  volume={25},
  number={1},
  pages={72--83},
  year={2000},
  publisher={IEEE}
}

@article{Delay,
  title={Delay and queue aware adaptive scheduling-based MAC protocol for underwater acoustic sensor networks},
  author={Zhuo, Xiaoxiao and Qu, Fengzhong and Yang, Hong and Wei, Yan and Wu, Yezhou and Li, Jianghui},
  journal={IEEE Access},
  volume={7},
  pages={56263--56275},
  year={2019},
  publisher={IEEE}
}

@article{aqua-sim-fg,
  author       = {Jiani Guo and
                  Shanshan Song and
                  Hao Chen and
                  Bingwen Huangfu and
                  Jun Liu and
                  Jun{-}Hong Cui},
  title        = {Aqua-Sim Fourth Generation: Toward General and Intelligent Simulation
                  for Underwater Acoustic Networks},
  journal      = {{IEEE} Internet Things J.},
  volume       = {12},
  number       = {15},
  pages        = {30203--30214},
  year         = {2025},
}

@INPROCEEDINGS{clock,
  author={Weng, Yang and Matsuda, Takumi and Sekimori, Yuki and Pajarinen, Joni and Peters, Jan and Maki, Toshihiro},
  booktitle={2022 IEEE/OES Autonomous Underwater Vehicles Symposium (AUV)}, 
  title={Time Synchronization Scheme of Underwater Platforms Using Wireless Acoustic and Optical Communication}, 
  year={2022},
  volume={},
  number={},
  pages={1-6},
}

@article{propagation-delay,
  title={Design of a propagation-delay-tolerant MAC protocol for underwater acoustic sensor networks},
  author={Guo, Xiaoxing and Frater, Michael R and Ryan, Michael J},
  journal={IEEE journal of oceanic engineering},
  volume={34},
  number={2},
  pages={170--180},
  year={2009},
  publisher={IEEE}
}

@inproceedings{sfama,
  title={Slotted FAMA: a MAC protocol for underwater acoustic networks},
  author={Molins, Marcal and Stojanovic, Milica},
  booktitle={OCEANS 2006-Asia Pacific},
  pages={1--7},
  year={2006},
  organization={IEEE}
}

@article{distance,
  title={Distance aware collision avoidance protocol for ad-hoc underwater acoustic sensor networks},
  author={Peleato, Borja and Stojanovic, Milica},
  journal={IEEE Communications letters},
  volume={11},
  number={12},
  pages={1025--1027},
  year={2007},
  publisher={IEEE}
}

@article{terminal,
  title={Handling triple hidden terminal problems for multichannel MAC in long-delay underwater sensor networks},
  author={Zhou, Zhong and Peng, Zheng and Cui, Jun-Hong and Jiang, Zaihan},
  journal={IEEE Transactions on Mobile Computing},
  volume={11},
  number={1},
  pages={139--154},
  year={2011},
  publisher={IEEE}
}

@article{propagation,
  title={Propagation and scattering effects in underwater acoustic communication channels},
  author={Van Walree, Paul A},
  journal={IEEE Journal of Oceanic Engineering},
  volume={38},
  number={4},
  pages={614--631},
  year={2013},
  publisher={IEEE}
}

@inproceedings{pmac,
  title={PMAC: a real-world case study of underwater MAC},
  author={Le, Son N and Zhu, Yibo and Peng, Zheng and Cui, Jun-Hong and Jiang, Zaihan},
  booktitle={Proceedings of the 8th International Conference on Underwater Networks \& Systems},
  pages={1--8},
  year={2013}
}

@article{uchannel,
  title={Underwater acoustic communication channels: Propagation models and statistical characterization},
  author={Stojanovic, Milica and Preisig, James},
  journal={IEEE communications magazine},
  volume={47},
  number={1},
  pages={84--89},
  year={2009},
  publisher={IEEE}
}

@inproceedings{towards,
  title={Towards achieving long-lifespan and self-sustained monitoring of coastal environments},
  author={Wang, Lei and Lei, Yu and Li, Baikun and Cui, Jun-Hong},
  booktitle={2014 IEEE International Conference on Systems, Man, and Cybernetics (SMC)},
  pages={3413--3418},
  year={2014},
  organization={IEEE}
}

@article{adhoc,
  title={Underwater acoustic sensor networks: research challenges},
  author={Akyildiz, Ian F and Pompili, Dario and Melodia, Tommaso},
  journal={Ad hoc networks},
  volume={3},
  number={3},
  pages={257--279},
  year={2005},
  publisher={Elsevier}
}
\vspace{-2em}
\begin{IEEEbiography}
  [{\includegraphics[width=1in,height=1.25in,clip,keepaspectratio]{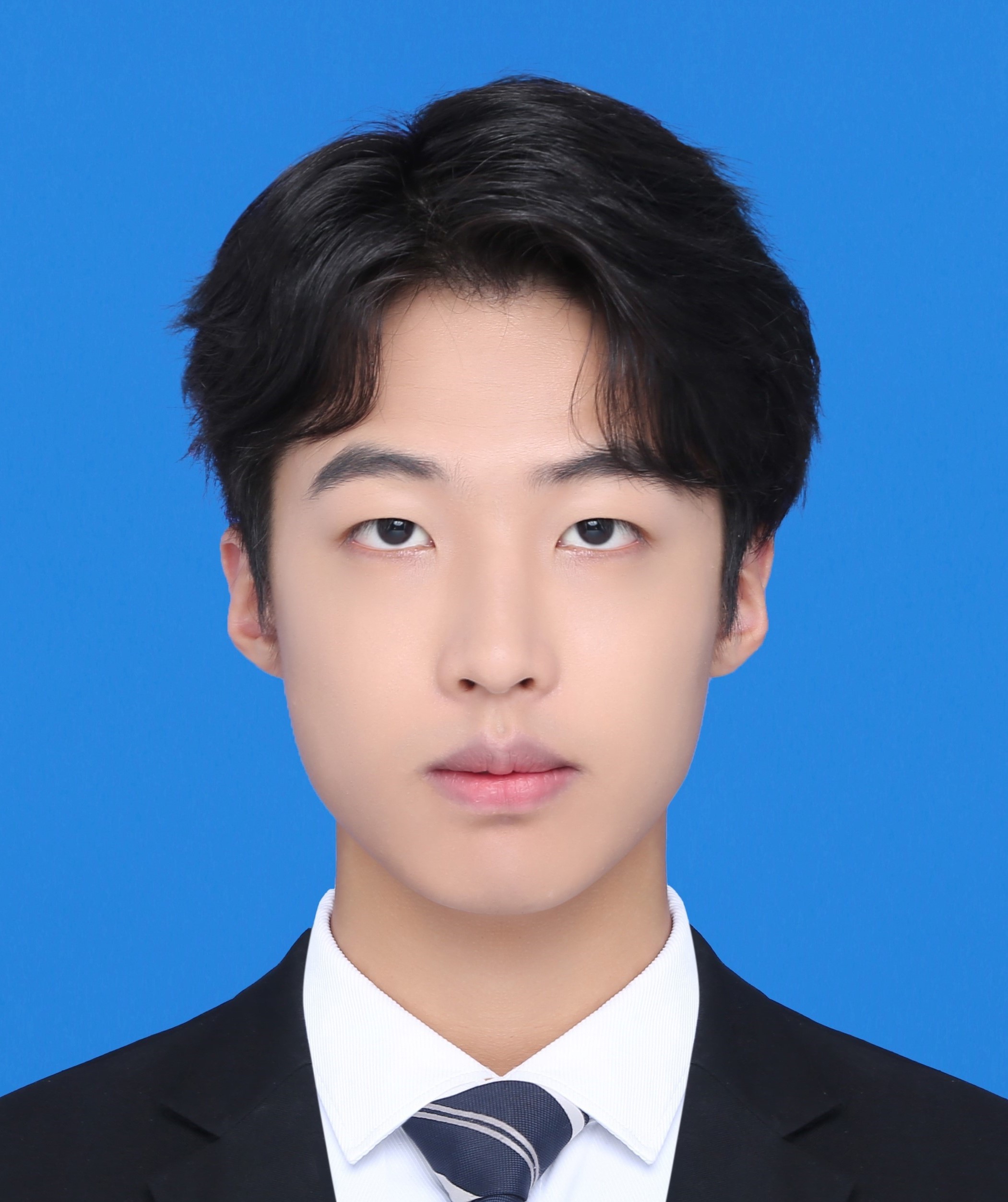}}]{Bingwen Huangfu} received the B.E. degree in computer science and technology from Jilin University, Changchun, China, in 2022. He is currently working toward the PhD degree at the College of Computer science and technology at Jilin University, Changchun, China.
  His current research interests include network architecture, resource allocation, and machine learning for underwater acoustic networks.
\end{IEEEbiography}

\begin{IEEEbiography}
  [{\includegraphics[width=1in,height=1.25in,clip,keepaspectratio]{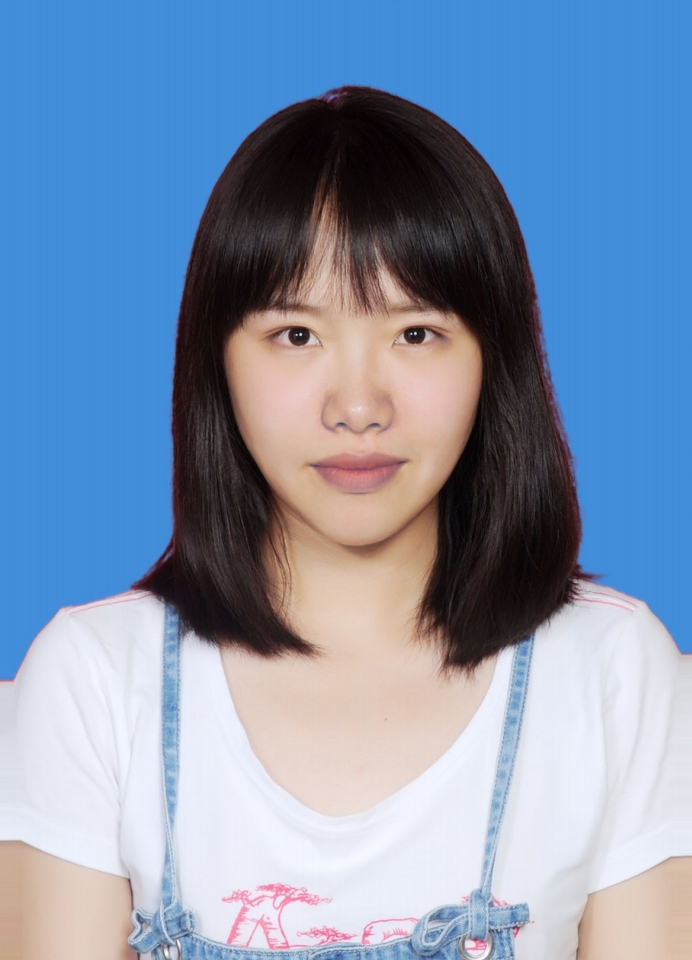}}]{Jiani Guo} received the BS degree (2016) in computer science and technology from Beijing Jiaotong University, Beijing, China, received PhD degree (2024) in Jilin University, Changchun, China. She is currently a Postdoctoral Researcher with the Department of Computer science and technology, Jilin University. Her current research interests include MAC protocols design and performance analysis for underwater acoustic networks.
\end{IEEEbiography}

\begin{IEEEbiography}
  [{\includegraphics[width=1in,height=1.25in,clip,keepaspectratio]{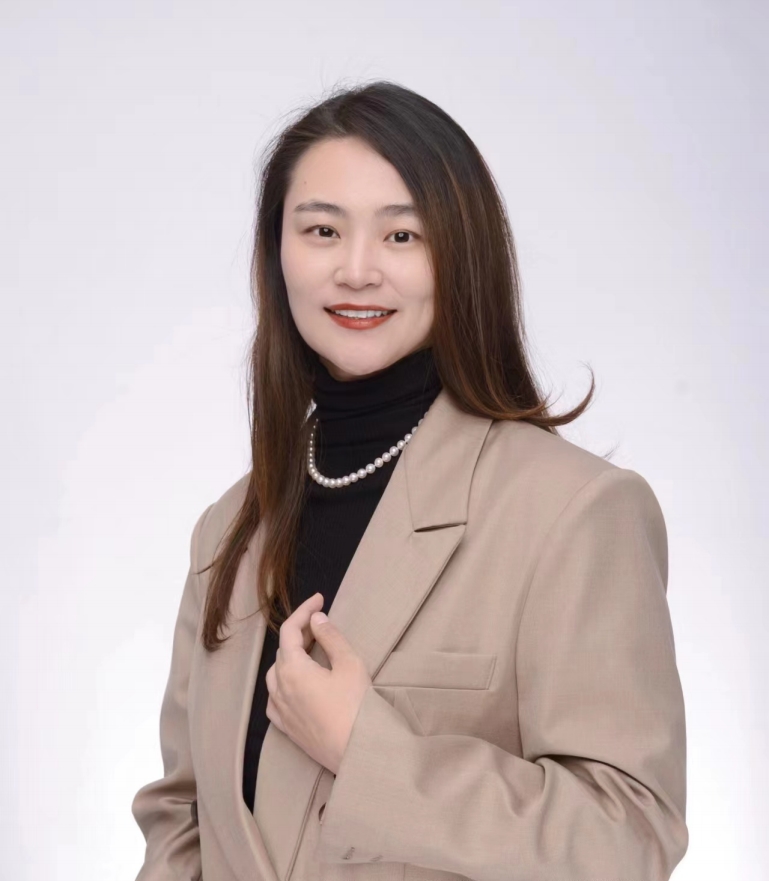}}]{Shanshan Song} (Member, IEEE) received the BS degree (2011) and MS degree (2014) in computer science and technology from Jilin University, China, received PhD degree (2018) in Management science and engineering from Jilin University, China. She was a Post-Doctoral Researcher with the Department of Computer science and technology, Jilin University, Changchun, China. She is currently an associate professor with the Department of Computer science and technology, Jilin University. Her major research focuses on underwater data collection, localization and navigation and machine learning. She serves as the WUWNet' 2023 Publication chair.
\end{IEEEbiography}

\begin{IEEEbiography}
  [{\includegraphics[width=1in,height=1.25in,clip,keepaspectratio]{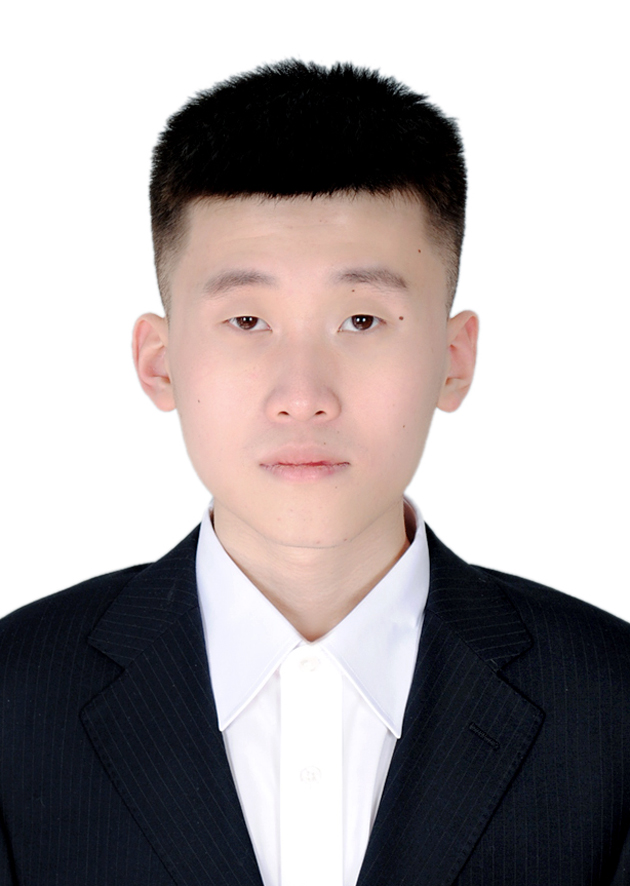}}]{Nan Sun} received the B.E. degree in software engineering from Yanshan University, Qinhuangdao, China, in 2023. He is currently working toward the M.S. degree with the College of Software Engineering, Jilin University, Changchun, China. His current research involves medium access control protocols for underwater acoustic networks.
\end{IEEEbiography}

\begin{IEEEbiography}
  [{\includegraphics[width=1in,height=1.25in,clip,keepaspectratio]{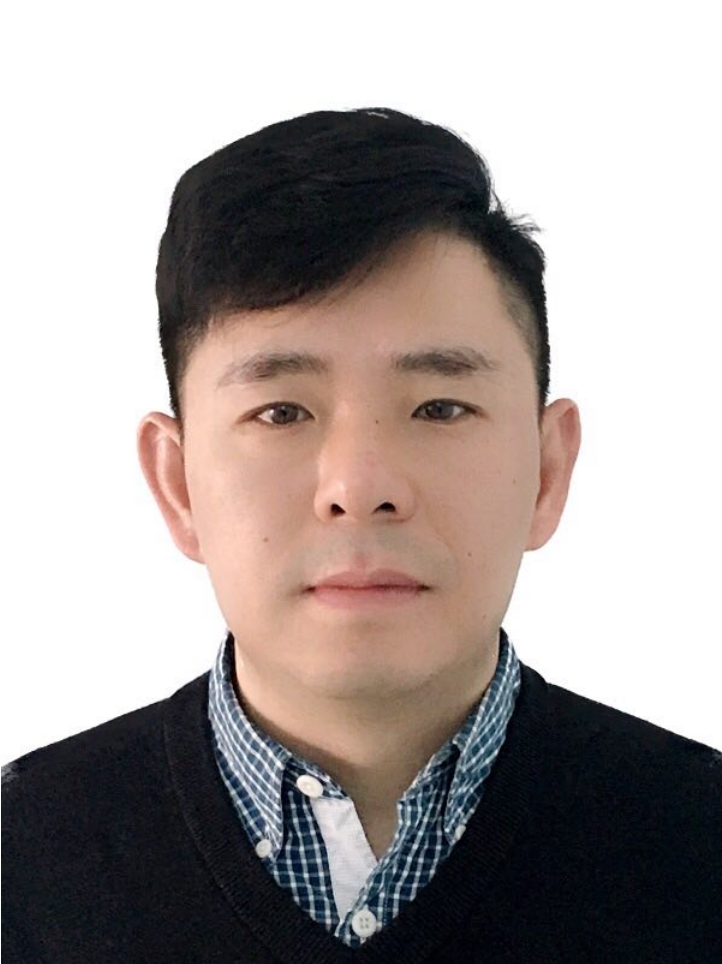}}]{Jun Liu} (Member, IEEE) received the BS degree (2002) in computer science from Wuhan University, China, the PhD degree (2013) in Computer Science and Engineering from University of Connecticut, USA. Currently, he is a professor of the School of Electronic and Information Engineering at Beihang University, Beijing, China, also a part-time professor of the Robotics Research Center, Peng Cheng Laboratory, Shenzhen, China. His major research focuses on underwater acoustic networking, time synchronization, localization, network deployment, and also interested in operating system, cross layer design. He is a member of the IEEE Computer Society.
\end{IEEEbiography}

\begin{IEEEbiography}
  [{\includegraphics[width=1in,height=1.25in,clip,keepaspectratio]{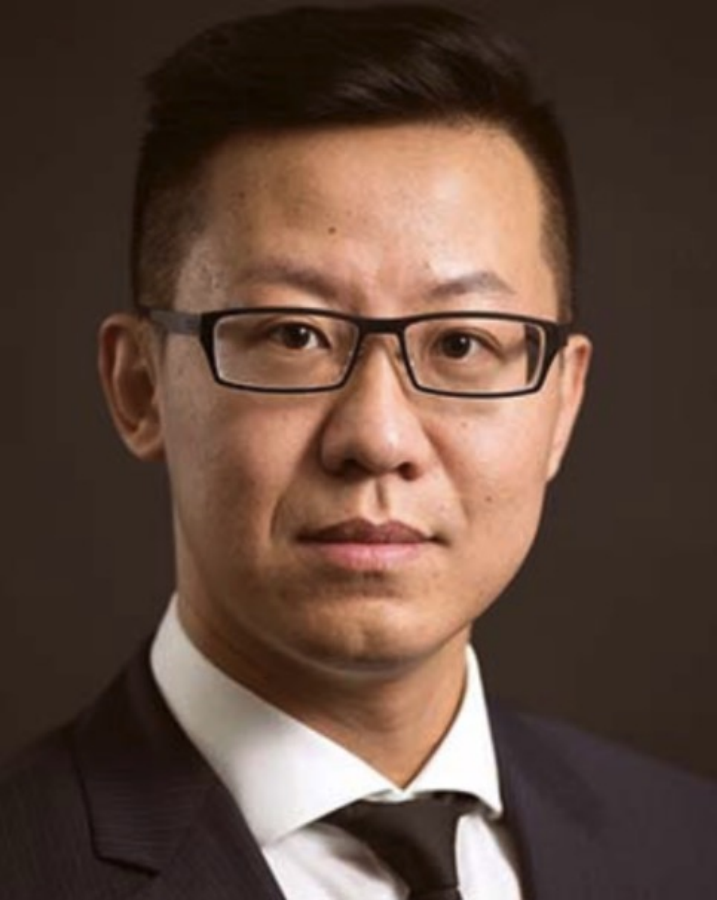}}]{Miao Pan} (Senior Member, IEEE) received the B.Sc. degree in electrical engineering from the Dalian University of Technology, Dalian, China, in 2004, the MASc. degree in electrical and computer engineering from the Beijing University of Posts and Telecommunications, Beijing, China, in 2007, and the Ph.D. degree in electrical and computer engineering from the University of Florida, Gainesville, FL, USA, in 2012. He is currently an Associate Professor with the Department of Electrical and Computer Engineering, University of Houston, Houston, TX, USA. His research interests include wireless/AI for AI/wireless, deep learning privacy, cybersecurity, and underwater communications and networking. He was the recipient of the NSF CAREER Award in 2014, IEEE TCGCC (Technical Committee on Green Communications and Computing) Best Conference Paper Awards 2019, and Best Paper Awards in ICC 2019, VTC 2018, Globecom 2017 and Globecom 2015, respectively. Dr. Pan is the Editor of IEEE OPEN JOURNAL OF VEHICULAR TECHNOLOGY, an Associate Editor for ACM Computing Surveys and IEEE INTERNET OF THINGS Journal (Area 5: Artificial Intelligence for IoT), and was an Associate Editor for IEEE INTERNET OF THINGS Journal (Area 4: Services, Applications, and Other Topics for IoT) from 2015 to 2018. He is also a Technical Organizing Committee for several conferences such as TPC Co-Chair for Mobiquitous 2019 and ACM WUWNet 2019.
\end{IEEEbiography}

\end{document}